\documentclass[article,  twocolumn,showpacs,preprintnumbers,amsmath,amssymb]{revtex4}

\usepackage{graphicx}% Include figure files
\usepackage{dcolumn}% Align table columns on decimal point
\usepackage{bm}% bold math
\numberwithin{equation}{section}

\begin{document}
\author{Xiaohua Wu}\author{Fei Wu}
\address{Department of Physics, Sichuan University, Chengdu 610064, China.}
\title{Designing  the optimal  quantum cloning machine for qubit case}

\begin{abstract}
Following the work of Niu and Griffiths, in \emph{Phys.Rev.A 58,
4377(1998)}, we shall investigate the problem, how to design the
optimal quantum cloning machines (QCMs) for qubit system, with the
help of  Bloch-sphere representation. In stead of the quality factor
there, the Fiur\'{a}\u{s}ek's optimal condition, where the optimal
cloning machine should maximize a convex mixture of the average
fidelity, is used as the optimality criterion in present protocol.
Almost all of the known optimal QCMs in previous works, the cloning
for states with fixed polar angle, the phase-covariant cloning, the
universal QCMs, the cloning for two arbitrary pure states, and the
mirror phase-covariant cloning, should be discussed in a systematic
way. The known results, the optimal fidelities for various input
ensembles according to different optimality criteria, are recovered
here. Our present scheme also offers a general way of constructing
the unitary transformation to realize the optimal cloning.
\end{abstract}


\maketitle

\section{introduction}
One of the fundamental no-go theorems in quantum mechanics is the
no-cloning theorem[1]. It is easy for us to make an arbitrary number
of copies of any types of information which arrive a classic
channel. However, if the information is encoded in terms of
nonorthogonal quantum states, to copy or clone the information which
arrives over a quantum channel is not possible without introducing
errors. Instead of obtaining  perfect copies, the idea of imperfect
cloning was introduced by Bu\u{z}ek and Hillery  who construct the
first \emph{quantum cloning machine } (QCM)[2]. Their work triggered
an explosion in the number of investigations on quantum cloning.

The fist QCM, which was introduced by Bu\u{z}ek and Hillery for
qubit case, is now known as the symmetric $1\rightarrow 2$ universal
quantum cloning machine
 (UQCM), a terms  comes from the fact that it copies equally well all the pure state.
The first study of non-universal or state-dependent symmetric
$1\rightarrow 2$ cloning, which is to clone at best two arbitrary
pure states of a qubit, has been given by Bru{\ss} \emph{et al.}[3].
The best-known example of state-dependent QCMs are the so-called
phase-covariant QCMs, which are defined as the QCMs that copy at
best states whose Bloch vector lies in the equator of the Bloch
sphere [4]. As an generalization of the phase-covariant cloning, the
problem of cloning to qubits, where the Euler angle $\theta$ is
specified and fixed, has been introduced by Kairimipour and
Rezakhani [5]. For this case, the optimal $1\rightarrow 2$ QCMs were
derived by Fiur\'{a}\u{s}ek  [6]. Recently, Bartkiewicz \emph{et
al.} provided an optimal cloner for the qubits with known $\sin
\theta$ [7]. On contrary to the symmetric $1\rightarrow 2$ QCMs,
where the outputs have the same fidelities, the asymmetric
$1\rightarrow 1+1$ QCMs could offer copies with different fidelity.
For qubit case, Niu and Griffiths derived, in particular, the
optimal asymmetric UQCM in their comprehensive study [8]. The same
result was found independently by Cerf who used an algebraic
approach [9,10]. A quantum circuit approach was pursued by
Bu\u{z}ek, Hillery and Bendik [11].

Our present work originates from the  open question, which has been
emphasized in the review article of Scarani \emph{et al.} [12], that
there is no general result concerning state-dependent cloning and
the zoology of cases \emph{is a priori} infinite. It is an
interesting task for us  to find a solution for this open problem.
Following the work of Niu and Griffiths in [8], we shall investigate
the problem, how to design the optimal quantum cloning machines
(QCMs) for qubit system, in the Bloch-sphere representation. In
stead of the quality factor there, the Fiur\'{a}\u{s}ek's optimal
condition, where the optimal  cloning machine should maximize a
convex mixture of the average fidelity, is used as the optimality
criterion in present protocol. Almost all of the known optimal QCMs
in previous works, the cloning for states with fixed polar angle,
the phase-covariant cloning, the universal QCMs, the cloning for two
arbitrary pure states, and the mirror phase-covariant cloning,
should be discussed  in a systematic way. The known results, the
optimal fidelities for various input ensembles according to
different optimality criteria, are recovered here. Our present
scheme also offers a general way of constructing the unitary
transformation to realize the optimal cloning.

Our present paper is organized as follows. Section II is a
preliminary section. The argument, how to choose the basis and
express the  Pauli operators according to it, is   given at the
beginning of this section.  After introducing the Bloch vector
transformation for the description of QCMs, we shall develop a
general scheme to  find out the optimal fidelities for various input
ensembles by applying the  Fiur\'{a}\u{s}ek's optimal condition.
  As the two non-centered asymmetric phase-independent QCMs, the cloning for states with fixed polar angle
and the phase-covariant cloning, should be discussed in Sec. III.
Two centered phase-independent examples, the universal QCMs and the
mirror phase-covariant cloning, can be found in  Sec. IV.  The
phase-dependent  problem,  cloning two arbitrary pure states, is
solved in Sec. V. Finally, some discussion about our results and a
short conclusion, will be given in the last section.

\section{preliminary}
\subsection{The Bloch-sphere representation}

We use $S$ to denote the set of pure qubit states to be cloned. For
each $\vert\psi\rangle\in S$, $\vert\psi\rangle =\alpha\vert
0\rangle+\beta\vert 1\rangle$ with
$\vert\alpha\vert^2+\vert\beta\vert^2=1$,  the distribution function
  $q(\alpha,\beta)$ should be known, $q(\alpha,\beta)>0$ and
  $\int_S q(\alpha,\beta)d\tau=1$ with $\int_S d\tau$ for measure.  At first,
we introduce a density matrix decided by $S$,
\begin{equation}
\rho_S=\int_S \vert\psi\rangle\langle\psi\vert q(\alpha,
\beta)d\tau,
\end{equation}
and calculate its eigenvalues and eigenvectors,
\begin{equation}
\rho_S\vert\uparrow\rangle=\frac{1+\lambda}{2}\vert\uparrow\rangle,
\rho_S\vert\downarrow\rangle=\frac{1-\lambda}{2}\vert\downarrow\rangle.
\end{equation}
After introducing the new basis with its vectors as $\vert
\uparrow\rangle$ and $\vert\downarrow\rangle$, we can define the
identity operator and Pauli matrices,
\begin{eqnarray}
\textbf{I}=\vert\uparrow\rangle\langle\uparrow\vert+\vert\downarrow\rangle\langle\downarrow\vert,
\sigma_x=\vert\uparrow\rangle\langle\downarrow\vert+\vert\downarrow\rangle\langle\uparrow\vert,\nonumber\\
\sigma_y=-i\vert\uparrow\rangle\langle\downarrow\vert+i\vert\downarrow\rangle\langle\uparrow\vert,
\sigma_z=\vert\uparrow\rangle\langle\uparrow\vert-\vert\downarrow\rangle\langle\downarrow\vert,
\end{eqnarray}
Now, every state (pure or mixed) $\rho$ for qubit case, can be
expressed by
\begin{equation}
\rho=\frac{1}{2}(\textbf{I}+\vec{\sigma}\cdot\vec{r}),
\end{equation}
where  $\vec{r}$ is a three-dimensional real vectors,
$\vec{r}=(r_x,r_y,r_z)$, with $r_i$ defined by
\begin{equation}
r_i=\texttt{Tr}(\sigma_i\rho),
\end{equation}
Specially,  the pure  state $\vert \psi\rangle$ is described  by the
unit vector $\vec{n}(\theta,\phi)\equiv(n_x,n_y,n_z)$,
\begin{equation}
n_x=\sin \theta\cos \phi, n_y=\sin \theta\sin \phi, n_z=\cos \theta,
\end{equation}
with $\theta$ and $\phi$  the polar and azimuthal angle in the Bloch
sphere. Notice $\sum_i n_i^2=1$. Considering the fact that
\begin{equation}
\vert\psi\rangle\langle\psi\vert=\frac{1}{2}(\textbf{I}+\vec{\sigma}\cdot
\vec{n})
 \end{equation}
and
\begin{equation}
\vert\psi\rangle=\cos \frac{\theta}{2}\vert\uparrow\rangle+\sin
\frac{\theta}{2}e^{-i\phi}\vert\downarrow\rangle,
\end{equation}
is equivalent with each other, we can rewrite $S$, the set of states
to be cloned, in the way like $S:\{\vec{n}(\theta,\phi),
q(\theta,\phi)\}$, where
 $\int_S q(\theta,\phi) d\tau=1$ with the
measure $\int_S d\tau=\frac{1}{4\pi}\int_0^{2\pi}\int_0^{\pi}\sin
\theta d\theta d\phi$.  Why the identity operator and Pauli
operators are defined by the basis
$\{\vert\uparrow\rangle,\vert\downarrow\rangle\}$ rather than the
original basis $\{\vert 0\rangle,\vert 1\rangle\}$? The reason is
that: according to Eqs. (2.3-5), the density matrix $\rho_S$ in  Eq.
(2.1) can be written as
\begin{equation}
\rho_S=\frac{1}{2}(\textbf{I}+\lambda\sigma_z)
\end{equation}
 where its  components, along the
$\hat{x}$ and $\hat{y}$ directions, take the value of zero,
\begin{equation}
\int_S\texttt{Tr}(\sigma_j\vert\psi\rangle\langle\psi\vert)
q(\theta,\phi) d\tau=0,
\end{equation}
where $j=x,y$.

Usually, we use $\rho^A$ and $\rho^B$ to denote the two copies of
cloning, and $\vec{r}^A$ and $\vec{r}^B$ for their corresponding
vectors in the Bloch-sphere representation,
\begin{equation}
\rho^k=\frac{1}{2}(\textbf{I}+\vec{\sigma}\cdot\vec{r}^k),
\end{equation}
where $k=A, B$. Taking $\vert\psi\rangle$ as the input for QCM,  the
\emph{single-copy fidelity } is found to be
\begin{equation} F^k_{\psi}=\frac{1}{2}(1+\vec{r}^k\cdot\vec{n})
\end{equation}
which just depends on the inner product of two vector $\vec{r}^k$
and $\vec{n}$. The \emph{average fidelity} can be defined as
\begin{equation}
\bar{F}^k=\int_S F^k_{\psi}q(\theta,\phi) d\tau. \end{equation}

Other parameters, which characterize  a given set of pure states,
are the so-called \emph{averaged length} $\overline{n_i}$\,
\begin{equation}
\overline{n_i }=\int_S n_i q(\theta,\phi) d\tau,
\end{equation}
and the \emph{fluctuation} $\overline{n_i^2}$,
\begin{equation}
 \overline{n^2_i
}=\int_S n^2_i q(\theta,\phi) d\tau,
\end{equation}
where the parameters $n_i $ with $i=x, y, z$, are defined in Eq.
(2.6). In general, $(\overline{n_i})^2\neq \overline{n_i^2}$. A
frequently used relation, $\sum_i \overline{n^2_i}=1$, can be
derived from the unit
 condition of $n_i$ in Eq. (2.6). Using Eq.
(2.10), one may get
\begin{equation}
\overline{n_x}=\overline{n_y}=0.
\end{equation}
In other words, by   properly choosing  the operators in Eq. (2.3),
we have only four parameters, $\overline{n_i^2}$ with $(i=x, y, z)$
and $\overline{n_z}$, for characterizing   $S$ to be cloned.

\subsection{Description of the QCMs in the Bloch-sphere representation}
Following the work of Niu and Griffiths [8], we shall describe the
QCMs in the Bloch-sphere representation.  We prepare the initial
state of the system with $\rho$ and denote the state of the
environment by $\rho_{\texttt{env}}$, the QCMs can be viewed as a
unitary transformation $U$ coupling $\rho$ and $\rho_{\texttt{env}}$
together,
\begin{equation}
 \rho\otimes \rho_{\texttt{env}}\rightarrow U\left(\rho \otimes
\rho_{\texttt{env}}\right)U^{\dagger}.\end{equation} The two final
copies of cloning, say, $\rho^A$ and $\rho^B$, can be get  by
performing the partial trace over the environment,
$\texttt{Tr}_{\texttt{env}}[U(\rho\otimes\rho_{\texttt{env}}U^{\dagger}]$.
Formally, $\rho^k=\sum_{m}E^k_m \rho  (E^k_m)^{\dagger}$ with
$\sum_m (E^k_m)^{\dagger}E^k_m=\textbf{I}$ [14,15].
 With $\vec{r}$ the Bloch vector for the input $\rho$ and $\vec{r}^k$ for the outputs $\rho^k$,
there exists  a map
\begin{equation} \vec{r}\rightarrow
\vec{r}^k=M^k \vec{r}+\vec{\delta}^k
\end{equation}
where $M^k$ is a $3\times 3$ real matrix, and $\vec{\delta}^k $ is a
constant vector. This is an\emph{ affine map,} mapping the Bloch
sphere into itself [14].

 As one of the main results of present work,
we shall design a series of QCMs with
\begin{equation}
\vec{r}^k=\left(
            \begin{array}{ccc}
              \eta_x^k & 0 & 0 \\
              0 & \eta_y^k & 0 \\
              0 & 0 & \eta_z^k \\
            \end{array}
          \right)\left(
                   \begin{array}{c}
                     r_x \\
                     r_y\\
                     r_z \\
                   \end{array}
                 \right)+\left(
                           \begin{array}{c}
                             0\\
                             0\\
                             \delta_z^k \\
                           \end{array}
                         \right)
                         \end{equation}
where both the matrices, $ M^A$ and $M^B$ in Eq. (2.18), are
diagonal \emph{at the same time}, while the shifts, $\delta^A_x$,
$\delta^A_y$, $\delta^B_x$,  and $\delta^B_y$, take a value of zero.
As it is proven in appendix A1, the  unitary transformation $U$,
which takes the forms in below equations,  is able to realize the
vector transformation defined in Eq. (2.19),
\begin{eqnarray}
U \vert\uparrow\rangle & \rightarrow &\cos \frac{\alpha}{2}\vert
u_+\rangle_{AB}\otimes\vert\uparrow\rangle_{C}
+\sin\frac{\alpha}{2}\vert
v_+\rangle_{AB}\otimes\vert\downarrow\rangle_C, \nonumber\\
U\vert\downarrow\rangle & \rightarrow
&\cos\frac{\tilde{{\alpha}}}{2}\vert u_-\rangle_{AB}\otimes\vert
\downarrow\rangle_{C}+\sin\frac{\tilde{\alpha}}{2}\vert
v_-\rangle_{AB}\otimes\vert\uparrow\rangle_C,\nonumber
\end{eqnarray}
where we suppose that the copies lie  in the two-dimensional space
$A$ and $B$ after the action of $U$ and the space $C$ for the output
state of ancilla.  One may check that $U\vert\uparrow\rangle$ and
$U\vert\downarrow\rangle$ are orthogonal under the condition that
$\langle u_{\pm}\vert v_{\pm}\rangle=0$,
\begin{eqnarray}
\vert u_+\rangle_{AB}&=&\cos \frac{\beta}{2}\vert
\uparrow\rangle_A\vert \uparrow\rangle_B+\sin \frac{\beta}{2}\vert
\downarrow\rangle_A\vert\downarrow\rangle_B,\nonumber\\
\vert u_-\rangle_{AB}&=&\sin \frac{\tilde{\beta}}{2}\vert
\uparrow\rangle_A\vert \uparrow\rangle_B+\cos
\frac{\tilde{\beta}}{2}\vert
\downarrow\rangle_A\vert\downarrow\rangle_B,\nonumber\\
\vert v_+\rangle_{AB}&=&\cos \frac{\gamma}{2}\vert
\uparrow\rangle_A\vert \downarrow\rangle_B+\sin
\frac{\gamma}{2}\vert
\downarrow\rangle_A\vert\uparrow\rangle_B,\nonumber\\
\vert v_-\rangle_{AB}&=&\sin \frac{\tilde{\gamma}}{2}\vert
\uparrow\rangle_A\vert \downarrow\rangle_B+\cos
\frac{\tilde{\gamma}}{2}\vert
\downarrow\rangle_A\vert\uparrow\rangle_B.
\end{eqnarray} We use
$\omega_i$ denote one of these free parameters,
$\omega_i\in\{\alpha,\tilde{\alpha},\beta,\tilde{\beta},\gamma,\tilde{\gamma}\}$.
According to the calculation, which is carefully done  in appendix
A1, the $U(\omega)$ in Eq. (2.20) is shown to be  consistent with
the Bloch vector transformation in Eq. (2.19),
\begin{eqnarray}
\eta^k_x&=&\cos \frac{\alpha}{2}\sin \frac{\tilde{\alpha}}{2}(\cos
\frac{\beta}{2}\cos \frac{\tilde{\gamma}}{2}^k+\sin \frac
{\beta}{2}\sin \frac{\tilde{\gamma}}{2}^k) \nonumber \\&&+ \sin
\frac{\alpha}{2}\cos \frac{\tilde{\alpha}}{2}(\cos
\frac{\tilde{\beta}}{2}\cos \frac{\gamma}{2}^k+\sin \frac
       {\tilde{\beta}}{2}\sin \frac{\gamma}{2}^k),
       \nonumber\\
       \eta^k_y&=&\cos
\frac{\alpha}{2}\sin \frac{\tilde{\alpha}}{2}(\cos
\frac{\beta}{2}\cos \frac{\tilde{\gamma}}{2}^k-\sin \frac
{\beta}{2}\sin \frac{\tilde{\gamma}}{2}^k) \nonumber \\&&+\sin
\frac{\alpha}{2}\cos \frac{\tilde{\alpha}}{2}(\cos
\frac{\tilde{\beta}}{2}\cos \frac{\gamma}{2}^k-\sin \frac
{\tilde{\beta}}{2}\sin \frac{\gamma}{2}^k),\nonumber\\
\eta_z^k&=&\frac{1}{2}(\cos^2 \frac{\alpha}{2}\cos
\beta+\sin^2\frac{\alpha}{2}\cos\gamma^k\nonumber\\
&&+\cos^2\frac{\tilde{\alpha}}{2}\cos\tilde{\beta}+\sin^2\frac{\tilde{\alpha}}{2}\cos
\tilde{\gamma}^k),\nonumber\\
\delta_z^k&=&\frac{1}{2}(\cos^2 \frac{\alpha}{2}\cos
\beta+\sin^2\frac{\alpha}{2}\cos\gamma^k\nonumber\\
&&-\cos^2\frac{\tilde{\alpha}}{2}\cos\tilde{\beta}-\sin^2\frac{\tilde{\alpha}}{2}\cos
\tilde{\gamma}^k),
\end{eqnarray}
where the the denotations,
\begin{eqnarray}
\gamma^A&=&\gamma,~\tilde{\gamma}^A=\tilde{\gamma}, \nonumber\\
\gamma^B&=&\pi-\gamma,~\tilde{\gamma}^B=\pi-\tilde{\gamma},
\end{eqnarray}
were used here for writing the results in appendix A1 into the
compact form of Eq. (2.21). \emph{It should be noted that the
$U(\omega)$ of Eq. (2.20) is defined by the basis vectors,
$\vert\uparrow\rangle$ and $ \vert\downarrow\rangle$, in Eq. (2.2)}.
For cloning the set of pure states $S$, by joining Eqs. (2.12-16)
and Eq. (2.19) together, the average fidelity in Eq. (2.13) should
be
\begin{equation} \bar{F}^k(\omega)=\frac{1}{2}(1+\eta^k_x
\overline{n_x^2}+\eta^k_y \overline{n_y^2}+\eta^k_z
\overline{n_z^2}+\delta^k_z \overline{n_z})
\end{equation}
with $k=A, B$. Both the averaged length $\overline{n_i}$ and
$\overline{n_i^2}$ are just  decided  by $S$, the set of states to
be cloned.

For a given case of cloning, the way of realizing the optimal
fidelity may be not unique [12]. Our general unitary transformation
in Eq. (2.20), as we shall show  with a series of examples, offers a
\emph{sufficient and systematic} way for designing the various kinds
of optimal QCMs for qubit system.
\subsection{The optimal QCMs}
 In present work, we shall show that the optimal QCMs, which have
been designed by different optimality criteria,  for examples, the
\emph{global fidelity} in [3], the \emph{quality factor} in [8], the
\emph{no-cloning inequality} in [9-11], \emph{etc.}, can also be
designed  by the  \emph{Fiur\'{a}\u{s}ek's optimal condition }[6,13]
where  the optimal asymmetric cloning machine should maximize a
convex mixture of the average  fidelity $\bar{F}^A$ and $\bar{F}^B$,
\begin{equation}
F(\omega)=p\bar{F}^A(\omega) +(1-p)\bar{F}^B(\omega),
\end{equation}
in which $p\in[0,1]$ is a parameter that controls the asymmetry of
the clone.  With the average fidelity in Eq. (2.23), considering the
fact that the averaged length $\overline{n_z}$ and
$\overline{n_i^2}$ have been decided by $S$, to design the optimal
QCMs is equivalent with finding out the the optimal  settings of
$\omega$ which satisfy the partial equations
\begin{equation}
\frac{\partial {F}(\omega)}{\partial \omega_j}\equiv p\frac{\partial
\bar{F}^A(\omega)}{\partial \omega_j}+(1-p)\frac{\partial
\bar{F}^B(\omega)}{\partial \omega_j}=0.
\end{equation}

 As an important case of Eq. (2.24), the optimal symmetric
cloning
 should  maximize the function,
 \begin{equation}
F(\omega)=\frac{1}{2}(\bar{F}^A(\omega)+\bar{F}^B(\omega)),
\end{equation}
where $p=1/2$ [6,13].  With  the denotations
$\eta_i=\frac{1}{2}(\eta_i^A+\eta_i^B)$
 and $\delta_z=\frac{1}{2}(\delta^A_z+\delta^B_z)$,
we express $F(\omega)$ in Eq. (2.26) with
$F(\omega)=\frac{1}{2}(1+\sum_i\eta_i\overline{n_i^2}+\delta_z
\overline{n_z})$ and prove  that the relations, $\frac{\partial
F}{\partial \gamma}\propto\sin (\frac{\pi}{4}-\frac{\gamma}{2})$ and
$\frac{\partial F}{\partial \tilde{\gamma}}\propto\sin
(\frac{\pi}{4}-\frac{\tilde{\gamma}}{2})$, always hold without
considering the actual values of $\overline{n^2_i}$ and
$\overline{n_z}$ for a given $S$ (see appendix A2).  In other words,
for the symmetric QCMs, the parameters
 $\gamma$ and $\tilde{\gamma}$ are fixed,
\begin{equation}
\gamma=\tilde{\gamma}=\frac{\pi}{2}.
\end{equation}
 Putting it back
into Eq. (2.21), we find $\eta_i^A=\eta_i^B$ and
$\delta^A_z=\delta^B_z$, the two copies now have the same fidelity,
$F_{\psi}^A=F^B_{\psi}$, as they should be. In conclusion, the
designing of symmetric QCMs can directly start from
\begin{equation}
F(\omega')=\frac{1}{2}(1+\sum_i\eta_i\overline{n_i^2}+\delta_z
\overline{n_z}).
\end{equation}
Under the condition in Eq. (2.27), we find the relations,
$\eta_i=\eta^A_i=\eta^B_i$ and $\delta_z=\delta^A_z=\delta^B_z$,
  with the  parameters
\begin{eqnarray}
\eta_x&=&\cos \frac{\alpha}{2}\sin \frac{\tilde{\alpha}}{2}\cos
(\frac{\pi}{4}-\frac{\beta}{2}) + \sin \frac{\alpha}{2}\cos
\frac{\tilde{\alpha}}{2}\cos
(\frac{\pi}{4}-\frac{\tilde{\beta}}{2}),    \nonumber\\
       \eta_y&=&\cos
\frac{\alpha}{2}\sin \frac{\tilde{\alpha}}{2} \cos (\frac{\pi}{4}+
\frac{\beta}{2}) +\sin \frac{\alpha}{2}\cos
\frac{\tilde{\alpha}}{2}\cos(\frac{\pi}{4}+
\frac{\tilde{\beta}}{2}),\nonumber\\
\eta_z&=&\frac{1}{2}(\cos^2 \frac{\alpha}{2}\cos
\beta+\cos^2\frac{\tilde{\alpha}}{2}\cos\tilde{\beta}),\nonumber\\
\delta_z&=&\frac{1}{2}(\cos^2 \frac{\alpha}{2}\cos
\beta-\cos^2\frac{\tilde{\alpha}}{2}\cos\tilde{\beta}).
\end{eqnarray}
There are four variables, $\omega'_j\in\{\alpha,
\tilde{\alpha},\beta,\tilde{\beta}\}$, left here, their optimal
settings should be decided by the equations $ \frac{\partial
F(\omega')}{\partial \omega'_j}=0 $.

\subsection{Classification  of the QCMS}
Here, we make a simple classification of the QCMs according to the
Bloch vector transformation in Eq. (2.19-21). Using  Eq. (2.19) and
Eq. (2.12), we have the single-copy fidelity,
\begin{eqnarray}
F^k_{\psi}&=&\frac{1}{2}(1+\eta^k_x \sin
^2\theta\cos^2\phi+\eta_y^k\sin^2\theta
\sin^2\phi\nonumber\\&&+\eta_z^k\cos^2\theta+\delta_z^k\cos \theta).
\end{eqnarray}
 A QCM is called \emph{phase-independent} if
$\eta^k_x=\eta^k_y$ because  that the single-copy fidelity is now
clearly independent of the phase $\phi$. Usually, the
phase-independent QCMs should appear in the cases with
$\overline{n_x^2}=\overline{n_y^2}$.
 A short discussion shall be applied here to show why this happens.
For the $S$ with $\overline{n_x^2}=\overline{n_y^2}$,    introducing
the denotation  $\eta_{\bot}^k=\frac{1}{2}(\eta_x^k+\eta_y^k)$, we
rewrite the average fidelity as
\begin{equation}
\bar{F}^k=\frac{1}{2}[1+\eta_{\bot}^k(1-\overline{n_z^2})+\eta_z^k\overline{n_z^2}+\delta^k_z\overline{n_z}].
\end{equation}
The expression of $\eta^k_{\bot}$ can be get from Eq. (2.21),
$\eta^k_{\bot}= \cos \frac{\alpha}{2}\sin
\frac{\tilde{\alpha}}{2}\cos \frac{\beta}{2}\cos
\frac{\tilde{\gamma}}{2}^k+ \sin \frac{\alpha}{2}\cos
\frac{\tilde{\alpha}}{2}\cos \frac{\tilde{\beta}}{2}\cos
\frac{\gamma}{2}^k$. From it,  $\partial \eta^k_{\bot}/\partial
\beta=-\frac{1}{2}\cos \frac{\alpha}{2}\sin
\frac{\tilde{\alpha}}{2}\sin \frac{\beta}{2}\cos
\frac{\tilde{\gamma}}{2}^k$ and $\partial \eta^k_{\bot}/\partial
\tilde{\beta}=-\frac{1}{2}\sin \frac{\alpha}{2}\cos
\frac{\tilde{\alpha}}{2}\sin \frac{\tilde{\beta}}{2}\cos
\frac{\gamma}{2}^k$. For the $\eta^k_z$ and $\delta^k_z$ defined in
Eq. (2.21), one may also get $\partial \eta^k_z/\partial
\beta=\partial\delta_z^k/\partial \beta=-\cos^2\frac{\alpha}{2}\sin
\beta$ and $\partial \eta^k_z/\partial
\tilde{\beta}=-\partial\delta_z^k/\partial
\tilde{\beta}=-\cos^2\frac{\tilde{\alpha}}{2}\sin \tilde{\beta}$. By
joining these results together, we find the average fidelity in Eq.
(2.31) with  $\frac{\partial \bar{F}^k}{\partial \beta}\propto\sin
\frac{\beta}{2}$ and $\frac{\partial \bar{F}^k}{\partial
\tilde{\beta}}\propto\sin \frac{\tilde{\beta}}{2}$. According to the
optimality equation in Eq. (2.25), the optimal settings of $\beta$
and $\tilde{\beta}$ can always be chosen as
\begin{equation}
\beta=\tilde{\beta}=0
\end{equation}
for cloning  the set of states with
$\overline{n_x^2}=\overline{n_y^2}$. Putting $\beta=\tilde{\beta}=0$
in Eq. (2.21), it can be seen that the transformation elements,
$\eta^k_x$ and $\eta^k_y$, now take a same expression, $\eta^k_x=
\eta^k_y\equiv\eta^k_{\bot}$. Other elements, $\delta^k_z$ and
$\eta^k_z$, can also be simplified by $\beta=\tilde{\beta}=0$,
\begin{eqnarray}
\eta_{\bot}^k&=&\cos \frac{\alpha}{2}\sin
\frac{\tilde{\alpha}}{2}\cos \frac{\tilde{\gamma}^k}{2}+\sin
\frac{\alpha}{2}\cos
\frac{\tilde{\alpha}}{2}\cos \frac{\gamma^k}{2},\nonumber\\
\eta_z^k&=&1-\sin^2\frac{\alpha}{2}\sin^2\frac{\gamma^k}{2}-\sin^2\frac{\tilde{\alpha}}{2}\sin
^2\frac{\tilde{\gamma}^k}{2},\nonumber\\
\delta_z^k&=&\sin^2\frac{\tilde{\alpha}}{2}\sin
^2\frac{\tilde{\gamma}^k}{2}-\sin^2\frac{\alpha}{2}\sin^2\frac{\gamma^k}{2}.
\end{eqnarray}
Putting   $\bar{F}^k$ in Eq. (2.31) with the above parameters back
to the optimal equation of Eq. (2.25), there are only four
parameters,
$\omega_j\in\{\alpha,\tilde{\alpha},\gamma,\tilde{\gamma}\}$, to be
decide there.

As we shall show later, almost all of the QCMs known yet are
phase-independent. Certainly, a QCM is \emph{phase-dependent} if
$\eta^k_x\neq \eta^k_y$. In short, a QCM is called \emph{centered
}if $\delta^k_z=0$, else, it is called \emph{non-centered}.

\begin{figure} \centering
\includegraphics[scale=0.45]{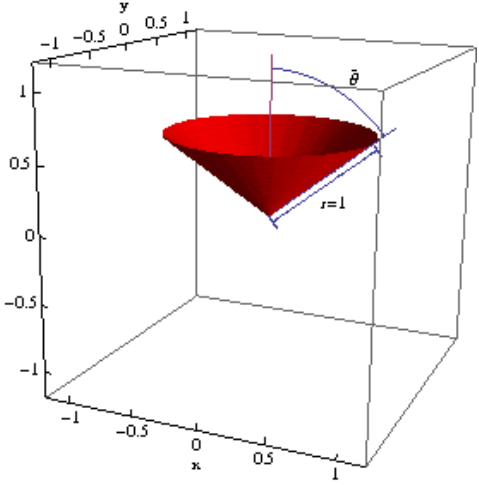}
\caption{\label{fig:epsart} The set of pure states
 with the Bloch vectors, $(\sin
\tilde{\theta}\cos \phi,\sin \tilde{\theta}\sin \phi, \cos
\tilde{\theta})$, where the polar angle has a given value
$\tilde{\theta}$ while the azimuthal angle is arbitrary.}
\end{figure}

\section{Non-centered phase-independent cloning}
\subsection{The set of sates with fixed polar angle}
Consider a   given set of states, $\vert\tilde{\psi}\rangle=\cos
\frac{\tilde{\theta}}{2}\vert
\uparrow\rangle+\sin\frac{\tilde{\theta}}{2}e^{-i\phi}\vert\downarrow\rangle$,
where the polar angle is fixed as $\tilde{\theta}$, $0\le
\tilde{\theta}\le\pi/2$,  while the azimuthal angle is arbitrary,
$\phi\in[0,2\pi]$. Its  Bloch representation  can be seen in FIG. 1.
A simple calculation shows
\begin{equation}
\overline{n_z}=\cos \tilde{\theta},
\overline{n_x^2}=\overline{n_y^2}=\frac{1}{2}\sin^2\tilde{\theta},\overline{n_z^2}=\cos^2\tilde{\theta}.
\end{equation}
The optimal symmetric cloning for this case has been given in [6].
Here,  its optimal asymmetric cloning is found with the settings
\begin{equation}
\beta=\tilde{\beta}=\alpha=0,\tilde{\alpha}=\pi
\end{equation}
 according to  the results in Appendix B1. The unitary
transformation $U$ takes the form,
\begin{eqnarray}
U\vert\uparrow\rangle&\rightarrow&
\vert\uparrow\uparrow\rangle\vert\uparrow\rangle,\nonumber\\
U\vert\downarrow\rangle&\rightarrow&(\sin\frac{\tilde{\gamma}}{2}\vert\uparrow\downarrow\rangle+
\cos
\frac{\tilde{\gamma}}{2}\vert\downarrow\uparrow\rangle)\vert\uparrow\rangle,
\end{eqnarray}
which comes from the general unitary transformation of Eq.(2.20)
with the  the optimal settings given at  Eq. (3.2). The
transformation matrix elements in Eq.(2.21) now become
\begin{equation}
\delta_z^k=\sin^2\frac{\tilde{\gamma}^k}{2},\eta^k_{\bot}=\cos\frac{\tilde{\gamma}^k}{2},\eta_z^k=\cos^2\frac{\tilde{\gamma}^k}{2}.
\end{equation}
The average fidelities in Eq. (2.31) are also known,
\begin{eqnarray}
\bar{F}^A=\frac{1}{2}(1+\cos\frac{\tilde{\gamma}}{2}\sin^2\tilde{\theta}+\cos^2\frac{\tilde{\gamma}}{2}
\cos^2\tilde{\theta}+\sin^2\frac{\tilde{\gamma}}{2}\cos\tilde{\theta}),\nonumber\\
\bar{F}^B=\frac{1}{2}(1+\sin\frac{\tilde{\gamma}}{2}\sin^2\tilde{\theta}+\sin^2\frac{\tilde{\gamma}}{2}
\cos^2\tilde{\theta}+\cos^2\frac{\tilde{\gamma}}{2}\cos\tilde{\theta}).\nonumber\\
\end{eqnarray}
The parameter$\tilde{\gamma}$  in above equations should be decided
by the asymmetric parameter $p$,
\begin{equation}
p=\frac{\cos \frac{\tilde{\gamma}}{2}\sin^2\tilde{\theta}+\sin
\tilde{\gamma}(\cos^2\tilde{\theta}-\cos \tilde{\theta})}{(\cos
\frac{\bar{\gamma}}{2}+\sin
\frac{\tilde{\gamma}}{2})\sin^2\tilde{\theta}+2\sin
\tilde{\gamma}(\cos^2\tilde{\theta}-\cos \tilde{\theta})},\nonumber
\end{equation}
according to the partial equation $p\partial \bar{F}^A/\partial
\tilde{\gamma}+(1-p)\partial \bar{F}^B/\partial \tilde{\gamma}=0$ in
Eq. (2.25). For the symmetric case, where $p=1/2$, the optimal
setting of $\tilde{\gamma}$ is  $\pi/2$ as it should be.

As an interesting result, it is found that the vector transformation
defined by Eq. (3.4) is just the  so-called \emph{amplitude damping}
(AD) known in [14]. Defining
$\varepsilon_{\texttt{NAD}}^k(\rho)=E^k_0\rho(E^k_0)^{\dagger}+E^k_1\rho(E^k_1)^{\dagger}$,
\begin{equation}
E_0^k=\left(
        \begin{array}{cc}
          1 & 0 \\
          0 & \cos\frac{\tilde{\gamma}^k}{2} \\
        \end{array}
      \right),E_1^k=\left(
                       \begin{array}{cc}
                         0 & \sin \frac{\tilde{\gamma}^k}{2} \\
                         0 & 0 \\
                       \end{array}
                     \right),
 \end{equation}
where
$\tilde{\gamma}^A=\tilde{\gamma},\tilde{\gamma}^B=\pi-\tilde{\gamma}$,
and $k=A, B$, we may verify that the two expressions, the vector
transformation  of Eq. (2.19) with its elements given by  Eq. (3.4)
and the amplitude damping defined above, are equivalent with each
other. The  effect of the amplitude damping on the Bloch sphere can
be seen in FIG. 2 [14].

\begin{figure} \centering
\includegraphics[scale=0.65]{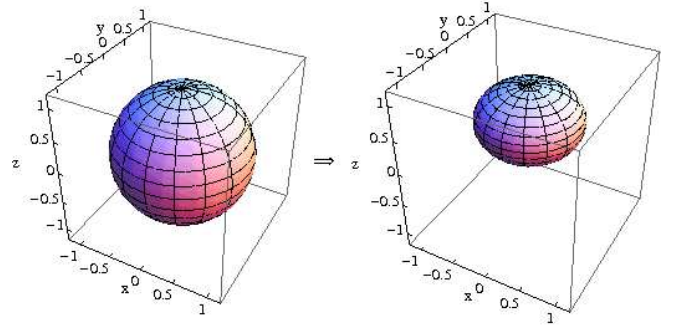}
\caption{\label{fig:epsart} The effect of the  amplitude damping on
the Bloch sphere, for
$\vec{r}\rightarrow\left(\frac{\sqrt{2}}{2}r_x,\frac{\sqrt{2}}{2}r_y,
\frac{1}{2}(1+r_z)\right)$. Note how the entire sphere shrinks to
the north-pole. It is the optimal symmetric QCM  for cloning the
states in FIG. 1.}
\end{figure}

Furthermore,  let   $\tilde{\gamma}=\pi/2$ according to $p=1/2$, we
have the optimal symmetric fidelity from Eq. (3.5),
\begin{equation}
F=\frac{1}{2}[1+\frac{\sqrt{2}}{2}\sin^2\tilde{\theta}+
\frac{1}{2}(\cos^2\tilde{\theta}+\vert\cos \tilde{\theta}\vert)],
\end{equation}
the result which has been given in [6].

\subsection{The phase-covariant cloning}
 The so-called phase-covariant QCMs were first introduced  in the problem of cloning the set of states, $\vert
\psi\rangle=\frac{\sqrt{2}}{2}(\vert\uparrow\rangle+e^{-i\phi}\vert\downarrow\rangle)$
with $\phi$ taking arbitrary values [4,16]. In present work, the
same problem   is concerned on following two aspects: the method of
deriving  it   by using
 the Fiur\'{a}\u{s}ek's optimal condition and its
 relationship with  the well-known \emph{generalized amplitude damping} (GAD) [14].

 The set of the states, which lie
in the equator of the Bloch sphere, are characterized by following
averaged parameters,
\begin{equation}
\overline{n_x^2}=\overline{n_y^2}=\frac{1}{2},\overline{n_z}=\overline{n_z^2}=0.
\end{equation}
Follow the argument in Appendix B2, the optimal settings for the
free parameters in Eq. (2.22) are chosen to be,
\begin{equation}
\beta=\tilde{\beta}=0,\tilde{\alpha}=\pi-\alpha,{\tilde{\gamma}}=\gamma,
\end{equation}
according to it,  the unitary transformation has the form,
\begin{eqnarray}
U\vert\uparrow\rangle
\rightarrow\cos\frac{\alpha}{2}\vert\uparrow\uparrow\uparrow\rangle+
\sin\frac{\alpha}{2}\left(\cos\frac{\gamma}{2}\vert\uparrow\downarrow\rangle+\sin
\frac{\gamma}{2}\vert\downarrow\uparrow\rangle\right)\vert\downarrow\rangle,\nonumber\\
U\vert\downarrow\rangle
\rightarrow\sin\frac{\alpha}{2}\vert\downarrow\downarrow\downarrow\rangle+
\cos\frac{\alpha}{2}\left(\sin\frac{\gamma}{2}\vert\uparrow\downarrow\rangle+\cos
\frac{\gamma}{2}\vert\downarrow\uparrow\rangle\right)\vert\uparrow\rangle,\nonumber
\end{eqnarray}
where $\alpha$ takes an arbitrary value. The transformation matrix
elements should be
\begin{equation}
\eta_{\bot}^k=\cos \frac{\gamma^k}{2},
\eta_z^k=\cos^2\frac{\gamma^k}{2}, \delta_z^k=\cos \alpha\sin^2
\frac{\gamma^k}{2}.
\end{equation}
Jointing it with Eq. (2.31), the average fidelity is found with
\begin{equation}
\bar{F}^A=\frac{1}{2}(1+\cos \frac{\gamma}{2}),
\bar{F}^B=\frac{1}{2}(1+\sin \frac{\gamma}{2}).
\end{equation}
One may also verify that $F^k_\psi=\bar{F^k}$. The result, which has
been derived by Niu and Griffiths  in [16], is  recovered here.
Certainly, the average fidelity in Eq. (3.11) can also be written in
the equivalent form where $p$ acts as the free variable. Applying
the optimal equation, $p\frac{\partial \bar{F}^A}{\partial
\gamma}+(1-p)\frac{\partial \bar{F}^B}{\partial \gamma}=0$, we find
\begin{equation}
\cos \frac{\gamma}{2}=\frac{p}{\sqrt{(1-p)^2+p^2}},\sin
\frac{\gamma}{2}=\frac{1-p}{\sqrt{(1-p)^2+p^2}},\nonumber
\end{equation} and
 get the fidelity,\begin{eqnarray}
\bar{F}^A=\frac{1}{2}(1+\frac{p}{\sqrt{(1-p)^2+p^2}}),\nonumber\\
\bar{F}^B=\frac{1}{2}(1+\frac{1-p}{\sqrt{(1-p)^2+p^2}}).\end{eqnarray}

Interestingly,  we find the vector transformation with its elements
in Eq. (3.10) to be the so-called \emph{generalized amplitude
damping} $\varepsilon^k_{\texttt{GAD}}$ with operator elements [14],
\begin{eqnarray}
E^k_0&=&\cos \frac{\alpha}{2}\left(
        \begin{array}{cc}
          1 & 0 \\
         0 & \cos \frac{\gamma^k}{2} \\
        \end{array}
      \right), E_1^k=\cos \frac{\alpha}{2}\left(
                                           \begin{array}{cc}
                                             0 & \sin \frac{\gamma^k}{2} \\
                                             0 & 0 \\
                                           \end{array}
                                         \right),\nonumber\\
E^k_2&=&\sin \frac{\alpha}{2}\left(
                             \begin{array}{cc}
                               \cos \frac{\gamma^k}{2} & 0 \\
                               0 & 1 \\
                             \end{array}
                           \right), E^k_3=\sin \frac{\alpha}{2}\left(
                                                                 \begin{array}{cc}
                                                                   0 & 0\\
                                                                   \sin \frac{\gamma^k}{2} & 0 \\
                                                                 \end{array}
                                                               \right),\nonumber\end{eqnarray}
where $\gamma^A=\gamma, \gamma^B=\pi-\gamma$. In FIG. 3, a vector
transformation,
$\vec{r}\rightarrow(\frac{\sqrt{2}}{2}r_x,\frac{\sqrt{2}}{2}r_y,
\frac{1}{4}+\frac{1}{2}r_z)$, is depicted  as an example for
$\varepsilon_{\texttt{GAD}}$.

\section{centered phase-independent cloning}
\subsection{The universal cloning}

A QCM is called \emph{universal} if it copies equally well all the
pure states $\vert\psi\rangle$ distributed in the surface of the
Bloch sphere with equal probability. This problem can be
characterized by
\begin{equation}
\overline{n_i^2}=\frac{1}{3}, \overline{n_z}=0
\end{equation}
with $i=x, y, z$. According to the calculation in appendix B3, we
find the optimal settings,
\begin{equation}
\tilde{\gamma}=\gamma, \beta=\tilde{\beta}=0, \tilde{\alpha}=\alpha,
\end{equation}
where $\alpha $ and $\gamma$ are decided by $p$,
\begin{eqnarray}
\cos \frac{\alpha}{2}=\frac{1}{\sqrt{2(1-p+p^2}},
\sin \frac{\alpha}{2}=\frac{\sqrt{1-2p+p^2}}{\sqrt{2(1-p+p^2)}}\nonumber\\
\cos \frac{\gamma}{2}=\frac{p}{\sqrt{2p^2-2p+1}}, \sin
\frac{\gamma}{2}=\frac{1-p}{\sqrt{2p^2-2p+1}},
\end{eqnarray}

The unitary transformation, with the above optimal settings, is
known
\begin{eqnarray}
 U\vert\uparrow\rangle
\rightarrow\cos\frac{\alpha}{2}\vert\uparrow\uparrow\uparrow\rangle+
\sin\frac{\alpha}{2}\left(\cos\frac{\gamma}{2}\vert\uparrow\downarrow\rangle+\sin
\frac{\gamma}{2}\vert\downarrow\uparrow\rangle\right)\vert\downarrow\rangle,\nonumber\\
U\vert\downarrow\rangle
\rightarrow\cos\frac{\alpha}{2}\vert\downarrow\downarrow\downarrow\rangle+
\sin\frac{\alpha}{2}\left(\sin\frac{\gamma}{2}\vert\uparrow\downarrow\rangle+\cos
\frac{\gamma}{2}\vert\downarrow\uparrow\rangle\right)\vert\uparrow\rangle.\nonumber
\end{eqnarray}
Its transformation elements should be
\begin{equation}
\eta^A_i=\frac{p}{{1-p+p^2}},\eta^B_i=\frac{1-p}{{1-p+p^2}},
\end{equation}
while $\delta^k_z=0$. From  Eq. (2.30), the single-copy fidelities
should be
\begin{equation}
F_{\psi}^A=\frac{1}{2}( 1+\frac{p}{{1-p+p^2}}),
F_{\psi}^B=\frac{1}{2}( 1+\frac{1-p}{{1-p+p^2}}),
\end{equation}
which saturate the no-cloning inequality
\begin{equation}
\sqrt{(1-F^A_{\psi})(1-F^B_{\psi})}\ge\frac{1}{2}-(1-F^A_{\psi})-(1-F^B_{\psi}).\nonumber
\end{equation}
Certainly, $\bar{F}^k=F^k_{\psi}$  since the equal probability for
each $\vert\psi\rangle$. The special transformation of Eq. (4.4), is
already  known to be the \emph{depolarizing channel }[8,10]

\begin{figure} \centering
\includegraphics[scale=0.7]{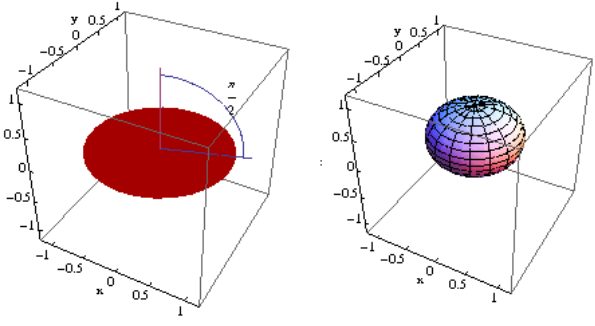}
\caption{\label{fig:epsart} On the left is the set of pure states in
the equator of the Bloch sphere. The generalized amplitude damping
in Eq. (3.4), which is the optimal QCM for  the states on left, is
chosen with
$\vec{r}\rightarrow(\frac{\sqrt{2}}{2}r_x,\frac{\sqrt{2}}{2}r_y,
\frac{1}{4}+\frac{1}{2}r_z)$. In fact, its center can arrange from
$(0,0,\frac{1}{2})$ to $(0,0,-\frac{1}{2})$.}
\end{figure}

\subsection{Centered symmetric phase-independent cloning}
For the set of states with
\begin{equation}
\overline{n^2_x}=\overline{n^2_y},\overline{n_z^2}>0,
\overline{n_z}=0,
\end{equation}
the optimal symmetric QCMs can be easily designed because there are
four parameters  already  decided,
\begin{equation}
\gamma=\tilde{\gamma}=\frac{\pi}{2}, \beta=\tilde{\beta}=0
\end{equation}
according to Eq. (2.27) and Eq. (2.32), respectively. As it is
calculated in appendix B4, the other two variables, $\alpha$ and
$\tilde{\alpha}$, should be
\begin{eqnarray}
 \tilde{\alpha}=\alpha,\nonumber\\\cos
 \alpha=\frac{\overline{n^2_z}}{\sqrt{(\overline{n_z^2})^2+2({1-\overline{n_z^2}})^2}},\nonumber\\
 \sin\alpha=\frac{\sqrt{2}(1-\overline{n_z^2})}{\sqrt{(\overline{n_z^2})^2+2(1-\overline{n_z^2})^2}}.
\end{eqnarray}
The unitary transformation with the optimal settings above is
\begin{eqnarray}
U\vert\uparrow\rangle
\rightarrow\cos\frac{\alpha}{2}\vert\uparrow\uparrow\uparrow\rangle+\frac
{\sqrt{2}}{2}
\sin\frac{\alpha}{2}\left(\vert\uparrow\downarrow\rangle+\vert\downarrow\uparrow\rangle\right)\vert\downarrow\rangle,\nonumber\\
U\vert\downarrow\rangle
\rightarrow\cos\frac{\alpha}{2}\vert\downarrow\downarrow\downarrow\rangle+\frac
{\sqrt{2}}{2}
\sin\frac{\alpha}{2}\left(\vert\uparrow\downarrow\rangle+\vert\downarrow\uparrow\rangle\right)\vert\uparrow\rangle,\nonumber
\end{eqnarray}
 Now, the diagonal elements of the transformation  matrix, which hold for both the copies, are
found to be,
\begin{equation}
\eta_{x}=\eta_y=\frac {\sqrt{2}}{2}\sin \alpha ,
\eta_z=\frac{1}{2}(1+\cos \alpha),
\end{equation} while
$\delta_z=0$.  Using Eq. (2.28), the optimal symmetric, which is for
the case defined in Eq. (4.6), is derived out,
\begin{equation}
F=\frac{1}{2}+\frac{1}{4}\left(\overline{n^2_z}+\sqrt{(\overline{n_z^2})^2+2(1-\overline{n_z^2})^2}\right),
\end{equation}

The  Bloch vector transformation in Eq. (4.9) can be verified to be
 equivalent with the so-called  \emph{symmetric Pauli channel}
(SP), $\varepsilon_{\texttt{SP}} (\rho)=\sum_mE_m \rho
E_m^{\dagger}$, with the operation elements,
\begin{eqnarray}
E_0=\sqrt{1-2a^2-b^2}\textbf{I},
E_1=a\sigma_x\nonumber\\
E_2=a\sigma_y, E_3=b\sigma_z, \end{eqnarray} with $a=\frac{1}{2}\sin
\frac {\alpha}{2}$ and
$b=\frac{\sqrt{2}}{2}(\cos\frac{\alpha}{2}-\frac{\sqrt{2}}{2}\sin
\frac{\alpha}{2})$.

 \emph{Example 1. The optimal universal symmetric
QCM}. Considering the special case, $\overline{n_i^2}=1/3$ and
$\overline{n_z}=0$, which appears in cloning all the unit Bloch
vectors with equal probability. With $\sin \frac
{\alpha}{2}=\sqrt{1/3}$ and $\cos \frac {\alpha}{2}=\sqrt{2/3}$ from
Eq. (4.8), the unitary transformation now takes the form
$U\vert\uparrow\rangle
\rightarrow\sqrt{2/3}\vert\uparrow\uparrow\rangle\vert\uparrow\rangle+
\sqrt{1/3}\vert\Psi^{+}\rangle\vert\downarrow\rangle$ and
$U\vert\downarrow\rangle
\rightarrow\sqrt{2/3}\vert\downarrow\downarrow\rangle\vert\downarrow\rangle+
\sqrt{1/3}\vert\Psi^{+}\rangle\vert\uparrow\rangle$ where
$\vert\Psi^{+}\rangle=\sqrt{1/2}(\vert\uparrow\downarrow\rangle+\vert\downarrow\uparrow\rangle).$
From Eq. (4.10), the optimal fidelity is $F=5/6$. The well-known
result given by B\v{u}zek and Hillery [2], is recovered here.
Certainly, the same result  can also be given by the asymmetric
universal QCM in Eq. (4.5) if $p$ is set with $1/2$ there. One may
verify that $a=b=\sqrt{1/6}$ from Eq. (4.11) and the symmetric Pauli
Channel  becomes the depolarizing channel.

\begin{figure} \centering
\includegraphics[scale=0.7]{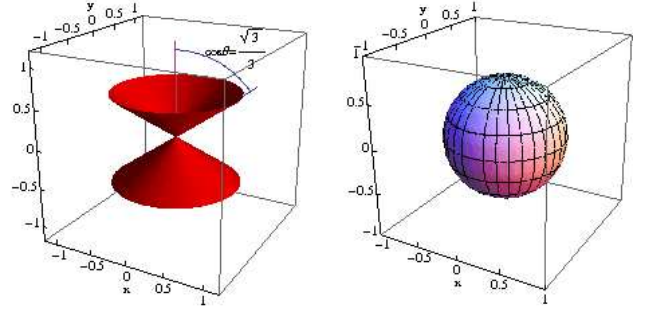}
\caption{\label{fig:epsart} On left is a  set  of states with $\sin
\theta=\sqrt{\frac{2}{3}}$ while $\phi$ taking an arbitrary value.
It is a special case of the so-called mirror phase-covariant cloning
with its optimal symmetric cloning machine, which is on the right,
to be $\vec{r}\rightarrow\frac{2}{3}\vec{r}$. All the Bloch vectors
shrink with a factor of $\frac{2}{3}$.}
\end{figure}

\emph{Example 2. Optimal mirror phase-covariant cloning.}  Recently,
Bartkiewicz $\sl{et~al.}$ proposed a quantum cloning machine, which
clones a qubit into two copies assuming known modulus of expectation
value of Pauli $\sigma_z$ matrix [7]. This is generalized version of
Fiur\'{a}\u{s}ek original one of cloning the set of states with
fixed value of $\theta$ [6]. The so-called \emph{mirror
phase-covariant cloning} (MPC) can be rephrased as to clone the
states $\vert\tilde{\psi}\rangle=\cos \frac{\tilde{\theta}}{2}\vert
\uparrow\rangle+\sin \frac{\tilde{\theta}}{2}e^{-i\phi}\vert
\downarrow\rangle$ with the polar angle takes one of the values,
$\tilde{\theta}$ and $\pi-\tilde{\theta}$, with an equal probability
while $\phi$ has an arbitrary value in the domain  $[0,2\pi]$. This
set of states is characterized by the averaged length and
fluctuations,
$\overline{n^2_x}=\overline{n^2_y}=\frac{1}{2}\sin^2\tilde{\theta},\overline{n_z^2}=\cos^2\tilde{\theta},
\overline{n_z}=0.$ With these values in hands, we have the optimal
setting, $\sin \alpha=\frac{\sqrt{2}\sin^2\tilde{\theta}}{\sqrt{\cos
^4\tilde{\theta}+2\sin ^4\tilde{\theta}}}$,$\cos
\alpha=\frac{\cos^2\tilde{\theta}}{\sqrt{\cos ^4\tilde{\theta}+2\sin
^4\tilde{\theta}}}$ from Eq. (4.8) and the optimal fidelity
\begin{equation}
F=\frac{1}{2}+\frac{1}{4}\left(\cos^2\tilde{\theta}+\sqrt{\cos^4
\tilde{\theta}+ 2\sin^4\tilde{\theta}}\right)\nonumber.
\end{equation} according to Eq. (4.10). An example of the mirror
phase-covariant with $\cos \theta=1/\sqrt{3}$ is depicted in Fig. 4
where $\vec{r}\rightarrow 2\vec{r}/3$.   This example  comes from
the symmetric Pauli channel in Eq. (4.11) with $a=b=1/\sqrt{6}$
there.

\section{Symmetric phase-dependent cloning}
In above sections, several types of phase-independent cloning have
been discussed. Usually, if $\eta^k_x\neq\eta^k_y$, the single-copy
fidelity  should depend on the actual value of $\phi$. Here, we
shall discuss  the symmetric phase-dependent QCMs for  case with
\begin{equation}
\overline{n_x^2}\neq \overline{n_y^2}=0.
\end{equation}
 As it is
calculated in appendix B5, the optimal setting for such case is
\begin{eqnarray}
\gamma=\tilde{\gamma}=\frac{\pi}{2},\alpha=0,\tilde{\alpha}=\pi,\nonumber\\
\sin(\frac{\pi}{4}-\frac{\beta}{2})=\frac{-\overline{n_x^2}+\sqrt{(\overline{n_x^2})^2+8(\overline{n_z^2}+\overline{n_z})^2}}{
4(\overline{n_z^2}+\overline{n_z})},
\end{eqnarray}
while $\tilde{\beta}$ takes an arbitrary value. This setting shall
simplify the unitary transformation in Eq. (2.20) into
\begin{eqnarray}
U\vert\uparrow\rangle\rightarrow(\cos
\frac{\beta}{2}\vert\uparrow\uparrow\rangle+\sin
\frac{\beta}{2}\vert\downarrow\downarrow\rangle)\vert\uparrow\rangle,\nonumber\\
U\vert\downarrow\rangle\rightarrow\sqrt{\frac{1}{2}}\left(\vert\uparrow\downarrow\rangle+\vert\downarrow\uparrow\rangle\right)\vert
\uparrow\rangle.
\end{eqnarray}
The vector transformation with its elements in Eq. (2.29) now take
the simple forms,
\begin{equation}
\eta_x=\cos(\frac{\pi}{4}-\frac{\beta}{2}),\eta_y=\cos(\frac{\pi}{4}+\frac{\beta}{2}),
\eta_z=\delta_z=\frac{1}{2}\cos \beta.
\end{equation}
Jointing Eq. (5.4) and Eq. (2.28) together, we find the optimal
symmetric fidelity
\begin{equation}
F=\frac{1}{2}\{1+\cos(\frac{\pi}{4}-\frac{\beta}{2})[\overline{n_x^2}+\sin(\frac{\pi}{4}-\frac{\beta}{2})
(\overline{n^2_z}+\overline{n_z})]\}.
\end{equation}
 In the operator-sum operation representation, the vector transformation with its elements in Eq. (5.4)
 is called the \emph{deformed  amplitude damping},
$\varepsilon_{\texttt{DAD}}$, with its elements as
\begin{equation}
E_0=\left(
        \begin{array}{cc}
          0 & \frac{\sqrt{2}}{2} \\
          \sin \frac{\beta}{2} & 0\\
        \end{array}
      \right),E_1=\left(
                      \begin{array}{cc}
                        \cos \frac{\beta}{2} & 0 \\
                        0 & \frac{\sqrt{2}}{2} \\
                      \end{array}
                    \right).
\end{equation}
The term DAD comes from the fact that $\varepsilon_{\texttt{DAD}}$
will reduce to the  amplitude damping in Eq. (3.6) if $\beta=0$.

\emph{Example 1. Cloning two arbitrary pure states.} The first study
of state-dependent cloning is to perform symmetric cloning of two
states [3], say,
$\vert\psi_1\rangle=\cos\frac{\tilde{\theta}}{2}\vert\uparrow\rangle+\sin\frac{\tilde{\theta}}{2}\vert\downarrow\rangle$
and
$\vert\psi_2\rangle=\cos\frac{\tilde{\theta}}{2}\vert\uparrow\rangle-\sin\frac{\tilde{\theta}}{2}\vert\downarrow\rangle$,
with equal probability $1/2$. These states lie in the $x-z$ plane of
the Bloch sphere.    Using $s$ to denote the overlap of the states,
$s=\cos\tilde{\theta}=\langle\psi_1\vert\psi_2\rangle$, with
$\overline{n_x^2}=1-s^2,\overline{n_y^2}=0,\overline{n_z^2}=s^2,\overline{n_z}=s,$
we find   the optimal setting of $\beta$ in Eq. (5.2), $\sin
(\frac{\pi}{4}-\frac{\beta}{2})=\frac{1}{4s}\left(-1+s+\sqrt{1-2s+9s^2}\right),$
and express the  fidelity in Eq. (5.5) in the way like
\begin{eqnarray}
F=\frac{1}{2}+\frac{\sqrt{2}}{32s}(1+s)(3-3s+\sqrt{1-2s+9s^2})\nonumber\\
\times\sqrt{-1+2s+3s^2+(1-s)\sqrt{1-2s+9s^2}}.\nonumber
\end{eqnarray}
It is the exact result which has been given in [3]. As an example,
the optimal symmetric cloning for two states with $s=\frac{1}{2}$ is
depicted in Fig. 5.

\begin{figure} \centering
\includegraphics[scale=0.6]{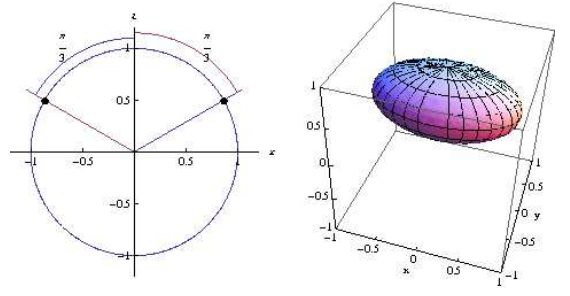}
\caption{\label{fig:epsart} On the left,  two unit vectors with the
relative angle  $\frac{2\pi}{3}$ are used to denote the two pure
states with their overlap to be $\frac{1}{2}$.
 The optimal symmetric QCM  is found  with the vector transformation, the so-called deformed amplitude
 damping, on the right where $\vec{r}\rightarrow\left(\frac{\sqrt{3}}{2}r_x,\frac{1}{2}r_y,\frac{\sqrt{3}}{4}(1+r_z)\right)$.
 Note that the rotation-invariance along the $\hat{z}$ axe, which  usually  appears in the phase-independent cloning, disappears  here.}
 \end{figure}

\emph{Example 2. Cloning two states with different probabilities.}
Consider  a set containing just  two states, $\vert\psi_1\rangle$
and $\vert\psi_2\rangle$,  with fixed overlap
$\langle\psi_1\vert\psi_2\rangle=\frac{1}{2}$. Suppose the
probability of $\vert\psi_1\rangle$ is denoted by $k$, $0\le k\le
\frac{1}{2}$, and $\vert\psi_2\rangle$ with a probability of $1-k$.
Following the argument in Sec. II, we chose a Bloch-sphere
representation where the Bloch vector of $\rho$,
$\rho=k\vert\psi_1\rangle\langle\psi_1\vert+(1-k)\vert\psi_2\rangle\langle\psi_2\vert$,
points along the direction of $\hat{z}$. The two states are
specified by their corresponding    Bloch vectors,
$\vec{r}_1=(\frac{\sqrt{3}(1-k)}{2\sqrt{1-3k+3k^2}},0,\frac{3k-1}{2\sqrt{1-3k+3k^2}})$
and
$\vec{r}_2=(\frac{-\sqrt{3}k}{2\sqrt{1-3k+3k^2}},0,\frac{2-3k}{2\sqrt{1-3k+3k^2}})$,
with the relative angle to be  $\frac{2\pi}{3}$. One may verify that
$\overline{n_z^2}=k r^2_{1z}+(1-k)r^2_{2z}$, $\overline{n_z}=k
r_{1z}+(1-k)r_{2z}$, $\overline{n_x^2}=k r^2_{1x}+(1-k)r^2_{2x}$ and
$\overline{n_y^2}=0$. The fidelities,  $F_{\psi_i}$ for each state
$\vert\psi_i\rangle$ and the average fidelity $\bar{F}$, are found
with $F_{\psi_i}=\frac{1}{2}\{1+\cos
(\frac{\pi}{4}-\frac{\beta}{2})[r_{ix}^2+\sin
(\frac{\pi}{4}-\frac{\beta}{2})(r_{iz}^2+r_{iz})]\}$ and
${F}=\frac{1}{2}\{1+\cos
(\frac{\pi}{4}-\frac{\beta}{2})[\overline{n_x^2}+\sin
(\frac{\pi}{4}-\frac{\beta}{2})(\overline{n_z^2}+\overline{n_z})]\}$.
 For the case with $k=\frac{1}{2}$, $\sin (\frac{\pi}{4}-\frac{\beta}{2})=\frac{1}{2}$, we find that
$F_{\psi_1}=F_{\psi_2}={F}=\frac{1}{2}(1+\frac{9\sqrt{3}}{16})$,
which  can also be derived from Eq. (5.9)  by letting
$s=\frac{1}{2}$ there. In general, if the probabilities $k$ and
$1-k$ are unequal, the state with the larger probability should
 have the higher fidelity to be cloned.
This can be seen from the extremal case with $ k\rightarrow 0$. With
 the optimal setting
$\sin(\frac{\pi}{4}-\frac{\beta}{2})\rightarrow\frac{\sqrt{2}}{2}$
get from Eq. (5.2).  The single-copy  fidelities,
$F_{\psi_1}\rightarrow\frac{7+3\sqrt{2}}{16}$ and
$F_{\psi_2}\rightarrow 1$, are different.  The average fidelity
${F}$, ${F}=kF_{\psi_1}+(1-k)F_{\psi_2}$, approaches 1.

\section{discussion}
The asymmetric phase-covariant cloning plays an important role in
the BB84 protocol [17-19]. In Sec. III,  the vector transformation
for such case is shown to be the well-known \emph{generalized
amplitude damping}. In principle, it can be detected by using the
approach of\emph{ quantum process tomography} [14] or the optimal
estimation scheme developed in [20].

In conclusion, with the help of the Bloch vector transformation, we
developed  a scheme to design the optimal QCMs according to the
Fiur\'{a}\v{s}ek's optimal condition. Our protocol is shown to be
successful in recovering the known optimal fidelities for various
input ensembles,  and it should represents a general solution for
the problem of optimal cloning in qubit system.

 \acknowledgments
We would like to thank Prof. Lu Xiaofu for discussion which helps us
a lot.

\begin{appendix}
\section{ Some proofs for section II}
\subsection{The proof for equation (2.21)}
The unitary transformation in Eq. (2.20), which  makes the
calculation of  the average fidelity in Eq. (2.23) with a simple
form, is rather special in the sense that both the matrices $M^A$
and $M^B$ are diagonal at the same time, a result which has not been
shown before. There are sufficient reasons for us to give the
derivation for Eq. (2.21)  in detail and to make sure that there is
no assumption needed here.

Our proof is given for the general mixed state rather than the pure
one. For an arbitrary state,
$\rho=\frac{1}{2}(\textbf{I}+\vec{\sigma}\cdot \vec{r})$   with $r$
for its length and $\vec{n}$ in Eq. (2.6) for its direction, we can
write it with an equivalent form,
$\rho=\frac{1+r}{2}\vert\psi\rangle\langle
\psi\vert+\frac{1-r}{2}\vert\psi^{\bot}\rangle\langle\psi^{\bot}\vert,$
with $\vert\psi\rangle=\cos\frac{\theta}{2}\vert\uparrow\rangle
+\sin \frac{\theta}{2}e^{-i\phi}\vert \downarrow\rangle$ and
$\vert\psi^{\bot}\rangle=\sin\frac{\theta}{2}\vert\uparrow\rangle
-\cos \frac{\theta}{2}e^{-i\phi}\vert \downarrow\rangle$. Using
$\vert\Psi\rangle$ and $\vert\Psi^{\bot}\rangle$ to denote
$U\vert\psi\rangle$ and $U\vert\psi^{\bot}\rangle$ respectively, we
shall do the calculations,
$\rho^A=\frac{1+r}{2}\texttt{Tr}_{BC}\vert\Psi\rangle\langle\Psi\vert)+
\frac{1-r}{2}\texttt{Tr}_{BC}\vert\Psi^{\bot}\rangle\langle\Psi^{\bot}\vert$
and
$\rho^B=\frac{1+r}{2}\texttt{Tr}_{AC}\vert\Psi\rangle\langle\Psi\vert)+
\frac{1-r}{2}\texttt{Tr}_{AC}\vert\Psi^{\bot}\rangle\langle\Psi^{\bot}\vert$.
After  writing  $\rho^k$ as
$\rho^k=\frac{1}{2}(\textbf{I}+\vec{\sigma}\cdot\vec{r}^k)$, we
shall prove that  the two vectors, $\vec{r}^k$ for the $k-th$ copy
and  $\vec{r}$ for input, should be related by the way in Eq.
(2.19).

Let's calculate $Tr_{BC}\vert\Psi\rangle\langle\Psi\vert$ at first.
According to the definition of $\vert\psi\rangle$ and the unitary
transformation in Eq. (2.20), we write $\vert\Psi\rangle$ as
\begin{widetext}
\begin{eqnarray}
\vert\Psi\rangle=(\cos \frac{\alpha}{2}\cos\frac{\beta}{2}\cos
\frac{\theta}{2}\vert\uparrow_A\rangle+\sin
\frac{\tilde{\alpha}}{2}\cos\frac{\tilde{\gamma}}{2}\sin
\frac{\theta}{2}
e^{-i\phi}\vert\downarrow_A\rangle)\otimes\vert\uparrow\rangle_B\rangle\otimes\vert\uparrow_C\rangle,\nonumber\\
+(\sin \frac{\tilde{\alpha}}{2}\sin \frac{\tilde{\gamma}}{2}\sin
\frac{\theta}{2}e^{-i\phi}\vert\uparrow_A\rangle+\cos
\frac{\alpha}{2}\sin \frac{\beta}{2}\cos
\frac{\theta}{2}\vert\downarrow_A\rangle)\otimes\vert\downarrow_B\rangle\otimes\vert
\uparrow_C\rangle,\nonumber\\
+(\cos \frac{\tilde{\alpha}}{2}\sin \frac{\tilde{\beta}}{2}\sin
\frac{\theta}{2}e^{-i\phi}\vert\uparrow_A\rangle+\sin
\frac{\alpha}{2}\sin \frac{\gamma}{2}\cos
\frac{\theta}{2}\vert\downarrow_A\rangle)\otimes\vert\uparrow_B\rangle\otimes\vert\downarrow_C\rangle,\nonumber\\
+(\sin \frac{\alpha}{2}\cos \frac{\gamma}{2}\cos
\frac{\theta}{2}\vert\uparrow_A\rangle+\cos
\frac{\tilde{\alpha}}{2}\cos \frac{\tilde{\beta}}{2}\sin
\frac{\theta}{2}e^{-i\phi}\vert\downarrow_A\rangle)\otimes\vert\downarrow_B\rangle\otimes\vert\downarrow_C\rangle.\nonumber
\end{eqnarray}
After performing the operation of partial trace, we shall get a
density matrix, $\left(
                   \begin{array}{cc}
                     a_{11} & a_{12} \\
                     a_{12}^* & a_{22} \\
                   \end{array}
                 \right)=
 \texttt{Tr}_{BC}\vert\Psi\rangle\langle
\Psi\vert$, with
\begin{eqnarray}
a_{11}=(\cos^2\frac{\alpha}{2}\cos^2\frac{\beta}{2}+\sin^2\frac{\alpha}{2}\cos^2\frac{\gamma}{2})\cos^2\frac{\theta}{2}
+(\sin^2\frac{\tilde{\alpha}}{2}\sin^2\frac{\tilde{\gamma}}{2}+\cos^2\frac{\tilde{\alpha}}{2}\sin^2\frac{\tilde{\beta}}{2})\sin^2\frac{\theta}{2},
\nonumber\\
a_{22}=(\cos^2\frac{\alpha}{2}\sin^2\frac{\beta}{2}+\sin^2\frac{\alpha}{2}\sin^2\frac{\gamma}{2})\cos^2\frac{\theta}{2}
+(\sin^2\frac{\tilde{\alpha}}{2}\cos^2\frac{\tilde{\gamma}}{2}+\cos^2\frac{\tilde{\alpha}}{2}\cos^2\frac{\tilde{\beta}}{2})\sin^2\frac{\theta}{2},
\nonumber\\
a_{12}=\frac{1}{2}\sin \theta[e^{i\phi}(\cos \frac{\alpha}{2}\sin
\frac{\tilde{\alpha}}{2}\cos \frac{\beta}{2}\cos
\frac{\tilde{\gamma}}{2}+\sin \frac{\alpha}{2}\cos
\frac{\tilde{\alpha}}{2}\cos \frac{\tilde{\beta}}{2}\cos
\frac{\gamma}{2}\nonumber\\
+e^{-i\phi}(\cos \frac{\alpha}{2}\sin \frac{\tilde{\alpha}}{2}\sin
\frac{\beta}{2}\sin \frac{\tilde{\gamma}}{2}+\sin
\frac{\alpha}{2}\cos \frac{\tilde{\alpha}}{2}\sin
\frac{\tilde{\beta}}{2}\sin \frac{\gamma}{2}].\nonumber
\end{eqnarray}
Noting  that the Bloch vector of $\vert \psi^{\bot}\rangle\langle
\psi^{\bot}\vert $ is $-\vec{n}$ and the vector for
$\vert\psi\rangle\langle\psi\vert$ is $\vec{n}$,
$\texttt{Tr}_{BC}\vert\Psi^{\bot}\rangle\langle\Psi^{\bot}\vert$ can
 be calculated by  substituting
 $\pi-\theta$ and $\pi+\phi$ for
 $\theta$ and $\phi$ in $\texttt{Tr}_{BC}\vert\Psi\rangle\langle\Psi\vert$, respectively.
Denoting $\rho^A=\left(
                             \begin{array}{cc}
                               \rho^A_{11} & \rho^A_{12} \\
                               \rho^A_{21} & \rho^B_{22} \\
                             \end{array}
                           \right)=
\frac{1+r}{2}\vert\Psi\rangle\langle
\Psi\vert+\frac{1-r}{2}\vert\Psi^{\bot}\rangle\langle\Psi^{\bot}\vert,$
the matrix elements should be,
\begin{eqnarray}
\rho_{11}^A=\frac{1+r\cos
\theta}{2}[\cos^2\frac{\alpha}{2}\cos^2\frac{\beta}{2}+\sin^2\frac{\alpha}{2}\cos^2\frac{\gamma}{2}]+\frac{1-r\cos
\theta}{2}[\sin^2\frac{\tilde{\alpha}}{2}\sin^2\frac{\tilde{\gamma}}{2}+\cos^2\frac{\tilde{\alpha}}{2}\sin^2\frac{\tilde{\beta}}{2}],\nonumber\\
\rho_{22}^A=\frac{1+r\cos
\theta}{2}[\cos^2\frac{\alpha}{2}\sin^2\frac{\beta}{2}+\sin^2\frac{\alpha}{2}\sin^2\frac{\gamma}{2}]+\frac{1-r\cos
\theta}{2}[\sin^2\frac{\tilde{\alpha}}{2}\cos^2\frac{\tilde{\gamma}}{2}+\cos^2\frac{\tilde{\alpha}}{2}\cos^2\frac{\tilde{\beta}}{2}],\nonumber
\end{eqnarray}
while $\rho^A_{12}=ra_{12}$ and $ \rho^A_{21}=(\rho_{12}^A)^*$.
Defining $\rho^
A=\frac{1}{2}(\textbf{I}+\vec{\sigma}\cdot\vec{r}^A)$ with
$r^A_i=\texttt{Tr}(\sigma_i\rho^A)$, we find $r_x^A=\eta_x^A r\sin
\theta\cos\phi$, $r_y^A=\eta^A_y r\sin \theta\sin\phi$, and
$r^A_z=\delta^A_z+\eta^A_z r\cos \theta$, where the parameters,
$\eta_i^A$ and $\delta^A_z$, take the form
\begin{eqnarray}
\eta_x^A=\cos \frac{\alpha}{2}\sin \frac{\tilde{\alpha}}{2}(\cos
\frac{\beta}{2}\cos \frac{\tilde{\gamma}}{2}+\sin
\frac{\beta}{2}\sin \frac{\tilde{\gamma}}{2})+ \sin
\frac{\alpha}{2}\cos \frac{\tilde{\alpha}}{2}(\cos
\frac{\tilde{\beta}}{2}\cos \frac{\gamma}{2}+\sin
\frac{\tilde{\beta}}{2}\sin \frac{\gamma}{2}),\nonumber\\
\eta_y^A=\cos \frac{\alpha}{2}\sin \frac{\tilde{\alpha}}{2}(\cos
\frac{\beta}{2}\cos \frac{\tilde{\gamma}}{2}-\sin
\frac{\beta}{2}\sin \frac{\tilde{\gamma}}{2})+\sin
\frac{\alpha}{2}\cos \frac{\tilde{\alpha}}{2}(\cos
\frac{\tilde{\beta}}{2}\cos \frac{\gamma}{2}-\sin
\frac{\tilde{\beta}}{2}\sin \frac{\gamma}{2}),\nonumber\\
\eta^A_z=\frac{1}{2}[\cos^2\frac{\alpha}{2}\cos
\beta+\sin^2\frac{\alpha}{2}\cos
\gamma+\cos^2\frac{\tilde{\alpha}}{2}\cos
\tilde{\beta}+\sin^2\frac{\tilde{\alpha}}{2}\cos
\tilde{\gamma}],\nonumber\\
\delta^A_z=\frac{1}{2}[\cos^2\frac{\alpha}{2}\cos
\beta+\sin^2\frac{\alpha}{2}\cos
\gamma-\cos^2\frac{\tilde{\alpha}}{2}\cos
\tilde{\beta}-\sin^2\frac{\tilde{\alpha}}{2}\cos
\tilde{\gamma}].\nonumber
\end{eqnarray}
Recalling the fact that the initial state $\rho$ is characterized by
the Bloch vector $\vec{r}=r(\sin \theta\cos \phi, \sin \theta\sin
\phi, \cos\theta)$, the vectors, $\vec{r}^A$ and $\vec{r}$, can be
proved to satisfy the vector transformation in Eq. (2.19). The
reduced density matrix $\rho^B$,
$\rho^B=\frac{1+r}{2}\texttt{Tr}_{AC}\vert\Psi\rangle\langle\Psi\vert+\frac{1-r}{2}\texttt{Tr}_{AC}\vert\Psi^{\bot}\rangle\langle\Psi^{\bot}\vert$,
can also be calculated in a similar way. The state
$U\vert\psi\rangle$ with the unitary transformation $U$ in Eq.
(2.20) can be rewritten as
\begin{eqnarray}
\vert\Psi\rangle=(\cos \frac{\alpha}{2}\cos \frac{\beta}{2}\cos
\frac{\theta}{2}\vert\uparrow_B\rangle+\sin
\frac{\tilde{\alpha}}{2}\sin \frac{\tilde{\gamma}}{2}\sin
\frac{\theta}{2}e^{-i\phi}\vert\downarrow_B\rangle)\otimes\vert\uparrow_A\rangle\otimes\vert\uparrow_C\rangle,\nonumber\\
=(\sin\frac{\tilde{\alpha}}{2}\cos \frac{\tilde{\gamma}}{2}\sin
\frac{\theta}{2}e^{-i\phi}\vert\uparrow_B\rangle+\cos
\frac{\alpha}{2}\sin \frac{\beta}{2}\cos
\frac{\theta}{2}\vert\downarrow_B\rangle)\otimes\vert\downarrow_A\rangle\otimes\vert\uparrow_C\rangle,\nonumber\\
=(\cos\frac{\tilde{\alpha}}{2}\sin \frac{\tilde{\beta}}{2}\sin
\frac{\theta}{2}e^{-i\phi}\vert\uparrow_B\rangle+\sin
\frac{\alpha}{2}\cos \frac{\gamma}{2}\cos
\frac{\theta}{2}\vert\downarrow_B\rangle)\otimes\vert\uparrow_A\rangle\otimes\vert\downarrow_C\rangle,\nonumber\\
=(\sin \frac{\alpha}{2}\sin \frac{\gamma}{2}\cos
\frac{\theta}{2}\vert\uparrow_B\rangle+\cos
\frac{\tilde{\alpha}}{2}\cos \frac{\tilde{\beta}}{2}\sin
\frac{\theta}{2}e^{-i\phi}\vert\downarrow_B\rangle)\otimes\vert\downarrow_A\rangle\otimes\vert\downarrow_C\rangle,\nonumber
\end{eqnarray}
 which is convenient for
performing $\texttt{Tr}_{AC}$. Let $ \left(
                               \begin{array}{cc}
                                 b_{11} & b_{12}\\
                                 b_{12}^* & b_{22} \\
                               \end{array}
                             \right)=\texttt{Tr}_{AC}\vert
                             \Psi\rangle\langle\Psi\vert$,  we shall
                             get\begin{eqnarray}
b_{11}=(\cos
^2\frac{\alpha}{2}\cos^2\frac{\beta}{2}+\sin^2\frac{\alpha}{2}\sin^2\frac{\gamma}{2})\cos^2\frac{\theta}{2}
+(\sin^2\frac{\tilde{\alpha}}{2}\cos
^2\frac{\tilde{\gamma}}{2}+\cos^2\frac{\tilde{\alpha}}{2}\sin^2\frac{\tilde{\beta}}{2})\sin^2\frac{\theta}{2},\nonumber\\
b_{22}=(\cos
^2\frac{\alpha}{2}\sin^2\frac{\beta}{2}+\sin^2\frac{\alpha}{2}\cos^2\frac{\gamma}{2})\cos^2\frac{\theta}{2}
+(\sin^2\frac{\tilde{\alpha}}{2}\sin
^2\frac{\tilde{\gamma}}{2}+\cos^2\frac{\tilde{\alpha}}{2}\cos^2\frac{\tilde{\beta}}{2})\sin^2\frac{\theta}{2},\nonumber\\
b_{12}=\frac{1}{2}\sin \theta[e^{i\phi}(\cos \frac{\alpha}{2}\sin
\frac{\tilde{\alpha}}{2}\cos \frac{\beta}{2}\sin
\frac{\tilde{\gamma}}{2}+\sin \frac{\alpha}{2}\cos
\frac{\tilde{\alpha}}{2}\cos \frac{\tilde{\beta}}{2}\sin
\frac{\gamma}{2})\nonumber\\
+e^{-i\phi}(\cos \frac{\alpha}{2}\sin \frac{\tilde{\alpha}}{2}\sin
\frac{\beta}{2}\cos \frac{\tilde{\gamma}}{2}+\sin
\frac{\alpha}{2}\cos \frac{\tilde{\alpha}}{2}\sin
\frac{\tilde{\beta}}{2}\cos \frac{\gamma}{2})].\nonumber
\end{eqnarray}
The reduced density matrix,
$\texttt{Tr}_{AC}\vert\Psi^{\bot}\rangle\langle\Psi^{\bot}\vert$,
can also be derived out by substituting $\pi-\theta$ and $\pi+\phi$
for the angles $\theta$ and $\phi$ in
$\texttt{Tr}_{AC}\vert\Psi\rangle\langle\Psi\vert$ given above.
Denote  $\vec{r}^B $ the Bloch vector for $ \rho^B$,
$\rho^B=\frac{1+r}{2}\texttt{Tr}_{AC}\vert\Psi\rangle\langle\Psi\vert+\frac{1-r}{2}
\texttt{Tr}_{AC}\vert\Psi^{\bot}\rangle\langle\Psi^{\bot}\vert$, and
$r^B_i=\texttt{Tr}({\sigma_i\rho^B})$, we shall get the results,
$r^B_x=\eta^B_xr\sin \theta\cos\phi$, $r_y^B=\eta^B_yr\sin
\theta\sin \phi$, and $r_z^B=\eta^B_zr\cos \theta+\delta^B_z$, where
\begin{eqnarray}
\eta_x^B=\cos \frac{\alpha}{2}\sin \frac{\tilde{\alpha}}{2}(\cos
\frac{\beta}{2}\sin \frac{\tilde{\gamma}}{2}+\sin
\frac{\beta}{2}\cos \frac{\tilde{\gamma}}{2})+ \sin
\frac{\alpha}{2}\cos \frac{\tilde{\alpha}}{2}(\cos
\frac{\tilde{\beta}}{2}\sin \frac{\gamma}{2}+\sin
\frac{\tilde{\beta}}{2}\cos \frac{\gamma}{2}),\nonumber\\
\eta_y^B=\cos \frac{\alpha}{2}\sin \frac{\tilde{\alpha}}{2}(\cos
\frac{\beta}{2}\sin \frac{\tilde{\gamma}}{2}-\sin
\frac{\beta}{2}\cos \frac{\tilde{\gamma}}{2})+\sin
\frac{\alpha}{2}\cos \frac{\tilde{\alpha}}{2}(\cos
\frac{\tilde{\beta}}{2}\sin \frac{\gamma}{2}-\sin
\frac{\tilde{\beta}}{2}\cos \frac{\gamma}{2}),\nonumber\\
\eta^B_z=\frac{1}{2}[\cos^2\frac{\alpha}{2}\cos
\beta-\sin^2\frac{\alpha}{2}\cos
\gamma+\cos^2\frac{\tilde{\alpha}}{2}\cos
\tilde{\beta}-\sin^2\frac{\tilde{\alpha}}{2}\cos
\tilde{\gamma}],\nonumber\\
\delta^B_z=\frac{1}{2}[\cos^2\frac{\alpha}{2}\cos
\beta-\sin^2\frac{\alpha}{2}\cos
\gamma-\cos^2\frac{\tilde{\alpha}}{2}\cos
\tilde{\beta}+\sin^2\frac{\tilde{\alpha}}{2}\cos
\tilde{\gamma}].\nonumber
\end{eqnarray}
With $\vec{r}=r(\sin \theta\cos \phi, \sin \theta\sin \phi, \cos
\theta)$, the two vectors, $\vec{r}^B$ and $\vec{r}$, are shown to
be related by the vector transformation  in Eq. (2.19). By
introducing the denotations, $\gamma^A=\gamma$,
$\tilde{\gamma}^A=\tilde{\gamma}$, $\gamma^B=\pi-\gamma$, and
$\tilde{\gamma}^B=\pi-\tilde{\gamma}$, all the transformation
elements derived here can be written into the compact form in  Eq.
(2.21).
\subsection{Proof for Eq. (2.27)}
According to the definition that
$\eta_i=\frac{1}{2}(\eta_i^A+\eta_i^B)$
 and $\delta_z=\frac{1}{2}(\delta^A_z+\delta^B_z)$, via Eq. (2.21),
 we have \begin{eqnarray}
 \eta_x=\cos \frac{\alpha}{2}\sin \frac{\tilde{\alpha}}{2}\cos (\frac{\pi}{4}
 -\frac{\beta}{2})\cos (\frac{\pi}{4}-\frac{\tilde{\gamma}}{2})+
 \sin \frac{\alpha}{2}\cos \frac{\tilde{\alpha}}{2}\cos (\frac{\pi}{4}-\frac{\tilde{\beta}}{2})\cos
 (\frac{\pi}{4}-\frac{\gamma}{2}),\nonumber\\
\eta_y=\cos \frac{\alpha}{2}\sin \frac{\tilde{\alpha}}{2}\cos
(\frac{\pi}{4}
 +\frac{\beta}{2})\cos (\frac{\pi}{4}-\frac{\tilde{\gamma}}{2})+
 \sin \frac{\alpha}{2}\cos \frac{\tilde{\alpha}}{2}\cos (\frac{\pi}{4}+\frac{\beta}{2})\cos
 (\frac{\pi}{4}-\frac{\gamma}{2}),\nonumber\\
\eta_z=\frac{1}{2}(\cos^2\frac{\alpha}{2}\cos \beta+
\cos^2\frac{\tilde{\alpha}}{2}\cos
\tilde{\beta}),\delta_z=\frac{1}{2}(\cos^2\frac{\alpha}{2}\cos
\beta- \cos^2\frac{\tilde{\alpha}}{2}\cos \tilde{\beta}).\nonumber
 \end{eqnarray}
One may easily verify that, $\frac{\partial
\eta_x}{\partial\gamma}\propto\sin(\frac{\pi}{4}-\frac{\gamma}{2})$,
$\frac{\partial
\eta_y}{\partial\gamma}\propto\sin(\frac{\pi}{4}-\frac{\gamma}{2})$,
and $\frac{\partial \eta_z}{\partial\gamma}=\frac{\partial
\delta}{\partial \gamma}=0$. For the fidelity in Eq. (2.26),
 $F(\omega)=\frac{1}{2}(1+\sum_i\eta_i\overline{n_i^2}+\delta_z
\overline{n_z})$, there should be  $\frac{\partial F}{\partial
\gamma}\propto\sin (\frac{\pi}{4}-\frac{\gamma}{2})$. In a similar
discussion, we find  $\frac{\partial F}{\partial
\tilde{\gamma}}\propto\sin
(\frac{\pi}{4}-\frac{\tilde{\gamma}}{2}).$ So, we can get
$\gamma=\tilde{\gamma}=\frac{\pi}{2}$ by letting $\frac{\partial
F}{\partial \gamma}=\frac{\partial F}{\partial \tilde{\gamma}}=0.$
\section{Optimal settings for  QCMs}
\subsection{The case where $\theta$ is fixed.}
From Eq. (2.31) and Eq. (3.1), we have the average fidelity,
$\bar{F}^k=\frac{1}{2}(1+\eta^k_{\bot}\sin^2\tilde{\theta}+\eta_z^k\cos^2\tilde{\theta}+\delta^k_z\cos
\tilde{\theta})$, with $\eta^k_{\bot}$, $\eta^k_z$ and $\delta_z^k$
given in Eq. (2.33). Using $\frac{\partial \bar{F}^A}{\partial
\gamma}=-\frac{1}{4}\sin
\frac{\alpha}{2}[\cos\frac{\tilde{\alpha}}{2}\sin\frac{\gamma}{2}\sin^2\tilde{\theta}+\sin\frac{\alpha}{2}\sin
\gamma(\cos^2\tilde{\theta}+\cos \tilde{\theta})]$ and
$\frac{\partial \bar{F}^B}{\partial \gamma}=\frac{1}{4}\sin
\frac{\alpha}{2}[\cos\frac{\tilde{\alpha}}{2}\cos\frac{\gamma}{2}\sin^2\tilde{\theta}+\sin\frac{\alpha}{2}\sin
\gamma(\cos^2\tilde{\theta}+\cos \tilde{\theta})]$,  with
$p\frac{\partial \bar{F}^A}{\partial \gamma}+(1-p)\frac{ \partial
\bar{F}^B}{\partial \gamma}=0$, the equation get from Eq. (2.25), we
arrive at  $\sin \frac{\alpha}{2}=0$. Putting it back into of
fidelity in Eq.(2.31), we have $\bar{F}^A=\frac{1}{2}[1+\sin
\frac{\tilde{\alpha}}{2}\cos
\frac{\tilde{\gamma}}{2}\sin^2\tilde{\theta}+\cos^2\tilde{\theta}
+\sin^2\frac{\tilde{\alpha}}{2}\sin^2\frac{\tilde{\gamma}}{2}(\cos
\tilde{\theta}-\cos^2\tilde{\theta})]$ and
$\bar{F}^B=\frac{1}{2}[1+\sin \frac{\tilde{\alpha}}{2}\sin
\frac{\tilde{\gamma}}{2}\sin^2\tilde{\theta}+\cos^2\tilde{\theta}
+\sin^2\frac{\tilde{\alpha}}{2}\cos^2\frac{\tilde{\gamma}}{2}(\cos
\tilde{\theta}-\cos^2\tilde{\theta})]$, where  there should be
$\frac{\partial\bar{F}^A}{\partial\tilde{\alpha}}=\frac{\cos
\frac{\tilde{\alpha}}{2}}{4}[\cos\frac{\tilde{\gamma}}{2}\sin^2\tilde{\theta}+2\sin\frac{\tilde{\alpha}}{2}\sin^2\frac{\tilde{\gamma}}{2}
(\cos \tilde{\theta}-\cos^2\tilde{\theta})]$ and
$\frac{\partial\bar{F}^B}{\partial\tilde{\alpha}}=\frac{\cos
\frac{\tilde{\alpha}}{2}}{4}[\sin\frac{\tilde{\gamma}}{2}\sin^2\tilde{\theta}+2\sin\frac{\tilde{\alpha}}{2}\cos^2\frac{\tilde{\gamma}}{2}
(\cos \tilde{\theta}-\cos^2\tilde{\theta})].$ One may easily verify
that the equation, $p\frac{\partial \bar{F}^A}{\partial
\tilde{\alpha}}+(1-p)\frac{ \partial \bar{F}^B}{\partial
\tilde{\alpha}}=0$, has a solution that $\cos
\frac{\tilde{\alpha}}{2}=0.$ Finally, we find $\alpha=0,
\tilde{\alpha}=\pi$,  the optimal setting which maximizes
$\bar{F}^k$ if $\cos \tilde{\theta}\ge0$.
\subsection{Optimal setting for the phase-covariant cloning}
For the phase-covariant case in Eq. (3.8), we also starts from Eq.
(2.31) and get the average fidelities, $\bar{F}^A=\frac{1}{2}(1+\sin
\frac{\alpha}{2}\sin \frac{\tilde{\alpha}}{2}\cos
\frac{\gamma}{2}+\sin \frac{\alpha}{2}\cos
\frac{\tilde{\alpha}}{2}\cos \frac{\gamma}{2})$ and
$\bar{F}^B=\frac{1}{2}(1+\sin \frac{\alpha}{2}\sin
\frac{\tilde{\alpha}}{2}\sin \frac{\gamma}{2}+\sin
\frac{\alpha}{2}\cos \frac{\tilde{\alpha}}{2}\sin
\frac{\gamma}{2})$. By requiring $\frac{\partial \bar{F}^A}{\partial
\gamma}\frac{\partial \bar{F}^B}{\partial
\tilde{\gamma}}-\frac{\partial \bar{F}^A}{\partial
\tilde{\gamma}}\frac{\partial \bar{F}^B}{\partial \gamma}=0$, we
shall find the setting $\gamma=\tilde{\gamma}$ which simplifies the
above average fidelities with $\bar{F}^A=\frac{1}{2}(1+\cos
\frac{\gamma}{2}\sin \frac{\alpha+\tilde{\alpha}}{2})$ and
$\bar{F}^B=\frac{1}{2}(1+\sin \frac{\gamma}{2}\sin
\frac{\alpha+\tilde{\alpha}}{2})$. Finally, the result,
$\alpha+\tilde{\alpha}=\pi$, can be easily achieved by asking
$\frac{\partial \bar{F}^A}{\partial \alpha}\frac{\partial
\bar{F}^B}{\partial \tilde{\alpha}}-\frac{\partial
\bar{F}^A}{\partial \tilde{\alpha}}\frac{\partial
\bar{F}^B}{\partial \alpha}=0$.

\subsection{Optimal  proof for asymmetric universal cloning}
 For the special case where
$\overline{n_i^2}=\frac{1}{3}$ while $\overline{n_i}=0$, the
fidelity in Eq. (2.31) should be,
\begin{eqnarray}
\bar{F}^A=\frac{1}{2}+\frac{1}{3}(\cos \frac{\alpha}{2}\sin
\frac{\tilde{\alpha}}{2}\cos \frac{\tilde{\gamma}}{2}+\sin
\frac{\alpha}{2}\cos \frac{\tilde{\alpha}}{2}\cos \frac{\gamma}{2})
+\frac{1}{6}(\cos^2\frac{\alpha}{2} +\sin^2\frac{\alpha}{2}\cos
\gamma +\cos^2\frac{\tilde{\alpha}}{2}
+\sin^2\frac{\tilde{\alpha}}{2}\cos \tilde{\gamma}),\nonumber\\
\bar{F}^B =\frac{1}{2}+\frac{1}{3}(\cos \frac{\alpha}{2}\sin
\frac{\tilde{\alpha}}{2}\sin \frac{\tilde{\gamma}}{2}+\sin
\frac{\alpha}{2}\cos \frac{\tilde{\alpha}}{2}\sin \frac{\gamma}{2})
+\frac{1}{6}(\cos^2\frac{\alpha}{2} -\sin^2\frac{\alpha}{2}\cos
\gamma +\cos^2\frac{\tilde{\alpha}}{2}
-\sin^2\frac{\tilde{\alpha}}{2}\cos \tilde{\gamma}),\nonumber
\end{eqnarray}
and a direct calculation  shows, $\frac{\partial \bar{F}^A}{\partial
\gamma}=-\frac{1}{6}\sin\frac{\alpha}{2}\sin\frac{\gamma}{2}(\cos\frac{\tilde{\alpha}}{2}
+2\sin\frac{\alpha}{2}\cos\frac{\gamma}{2})$, $\frac{\partial
\bar{F}^A}{\partial
\gamma}=\frac{1}{6}\cos\frac{\alpha}{2}\sin\frac{\gamma}{2}(\cos\frac{\tilde{\alpha}}{2}
+2\sin\frac{\alpha}{2}\sin\frac{\gamma}{2})$, $\frac{\partial
\bar{F}^A}{\partial
\tilde{\gamma}}=-\frac{1}{6}\sin\frac{\tilde{\alpha}}{2}\sin\frac{\tilde{\gamma}}{2}(\cos\frac{{\alpha}}{2}
+2\sin\frac{\tilde{\alpha}}{2}\cos\frac{\tilde{\gamma}}{2})$,
$\frac{\partial \bar{F}^A}{\partial
\tilde{\gamma}}=\frac{1}{6}\cos\frac{\tilde{\alpha}}{2}\sin\frac{\tilde{\gamma}}{2}(\cos\frac{{\alpha}}{2}
+2\sin\frac{\tilde{\alpha}}{2}\sin\frac{\tilde{\gamma}}{2})$,
$\frac{\partial \bar{F}^A}{\partial
\tilde{\alpha}}=\frac{1}{6}(\cos\frac{\alpha}{2}\cos\frac{\tilde{\alpha}}{2}\cos\frac{\tilde{\gamma}}{2}
-\sin\frac{\alpha}{2}\sin\frac{\tilde{\alpha}}{2}\cos\frac{\gamma}{2})
-\frac{1}{6}\sin \tilde{\alpha}(1-\cos \tilde{\gamma})$,
$\frac{\partial \bar{F}^B}{\partial
\tilde{\alpha}}=\frac{1}{6}(\cos\frac{\alpha}{2}\cos\frac{\tilde{\alpha}}{2}\sin\frac{\tilde{\gamma}}{2}
-\sin\frac{\alpha}{2}\sin\frac{\tilde{\alpha}}{2}\sin\frac{\gamma}{2})
-\frac{1}{6}\sin \tilde{\alpha}(1+\cos \tilde{\gamma})$,
$\frac{\partial\bar{F}^A}{\partial
\alpha}=\frac{1}{6}(\cos\frac{\alpha}{2}\cos\frac{\tilde{\alpha}}{2}\cos\frac{\gamma}{2}
-\sin\frac{\alpha}{2}\sin\frac{\tilde{\alpha}}{2}\cos\frac{\tilde{\gamma}}{2})
-\frac{1}{6}\sin\alpha(1-\cos\gamma)$, and
$\frac{\partial\bar{F}^B}{\partial
\alpha}=\frac{1}{6}(\cos\frac{\alpha}{2}\cos\frac{\tilde{\alpha}}{2}\sin\frac{\gamma}{2}
-\sin\frac{\alpha}{2}\sin\frac{\tilde{\alpha}}{2}\sin\frac{\tilde{\gamma}}{2})-\frac{1}{6}\sin\alpha(1+\cos\gamma)$.
With  $\frac{\partial \bar{F}^A}{\partial
\gamma}\frac{\partial\bar{F}^B}{\partial \tilde{\gamma}}
-\frac{\partial\bar{F}^A}{\partial
\tilde{\gamma}}\frac{\partial\bar{F}^B}{\partial \gamma}=0$ and
 $\frac{\partial
\bar{F}^A}{\partial \alpha}\frac{\partial\bar{F}^B}{\partial
\tilde{\alpha}} -\frac{\partial\bar{F}^A}{\partial
\tilde{\alpha}}\frac{\partial\bar{F}^B}{\partial \alpha}=0$, the two
equations come from Eq. (2.25), we find $\tilde{\alpha}=\alpha$ and
$\tilde{\gamma}=\gamma$ and arrive at the fidelities,
$\bar{F}^A=\frac{1}{2}+\frac{2}{3}\sin \alpha \cos \frac{\gamma}{2}
+\frac{1}{3}(\cos^2\frac{\alpha}{2} +\sin^2\frac{\alpha}{2}\cos
\gamma )$ and $\bar{F}^B=\frac{1}{2}+\frac{2}{3}\sin \alpha \sin
\frac{\gamma}{2} +\frac{1}{3}(\cos^2\frac{\alpha}{2}
-\sin^2\frac{\alpha}{2}\cos \gamma)$, containing just two
parameters, $\alpha$ and $\gamma$, here. Finally, from the  two
optimal equations, $p\frac{\partial \bar{F}^A}{\partial
\gamma}+(1-p)\frac{\partial \bar{F}^B}{\partial \gamma}=0$ and
$p\frac{\partial \bar{F}^A}{\partial \alpha}-(1-p)\frac{\partial
\bar{F}^B}{\partial \alpha}=0$, and the average fidelity above, we
may get the optimal settings of $\alpha$ and $\gamma$ in Eq.(4.3).

\subsection{Proof for Eq. (4.8)}
By letting $\gamma=\tilde{\gamma}=0$ for Eq. (2.29), we get the
fidelity, which is for the case defined in Eq. (4.6), with the
expression that $F=\frac{1}{2}+\frac{1}{4}\cos^2 \tilde{\theta}(\cos
\frac{\alpha}{2}+\cos^2\frac{\tilde{\alpha}}{2})+\frac{\sqrt{2}}{4}\sin^2\tilde{\theta}
(\cos \frac{\alpha}{2}\sin \frac{\tilde{\alpha}}{2}+\sin
\frac{\alpha}{2}\cos \frac{\tilde{\alpha}}{2})$. Requiring
$\frac{\partial F}{\partial \alpha}=\frac{\partial F}{\partial
\tilde{\alpha}}=0$, we shall  get the optimal settings of
$\tilde{\alpha}$ and  $\alpha$ in Eq. (4.8).
\subsection{Optimal setting for  Eq. (5.2)}
From Eq. (2.28) and Eq. (5.1 ), the fidelity should be
$F=\frac{1}{2}\{1+(1-\overline{{n^2_z}})[\cos \frac{\alpha}{2}\sin
\frac{\tilde{\alpha}}{2}\cos (\frac{\pi}{4}-\frac{\beta}{2})+\sin
\frac{\alpha}{2}\cos \frac{\tilde{\alpha}}{2}\cos
(\frac{\pi}{4}-\frac{\tilde{\beta}}{2})]+\frac{1}{2}\overline{{n_z}^2}(\cos^2\frac{\alpha}{2}\cos
\beta+\cos^2\frac{\tilde{\alpha}}{2}\cos
\tilde{\beta})+\frac{1}{2}{\overline{n_z}}(\cos^2\frac{\alpha}{2}\cos
\beta-\cos^2\frac{\tilde{\alpha}}{2}\cos \tilde{\beta})\}$. The
setting $\tilde{\alpha}=\pi$, which is get from the result
${\partial {F}}/{\partial \tilde{\beta}}\varpropto
\cos\frac{\tilde{\alpha}}{2}$, simplifies the fidelity with
$F=\frac{1}{2}\{1+(1-\overline{{n^2_z}})\cos \frac{\alpha}{2}\cos
(\frac{\pi}{4}-\frac{\beta}{2})+\frac{1}{2}(\overline{{n_z}^2}+\overline{n_z})\cos^2\frac{\alpha}{2}\cos
\beta\}$. The setting $\alpha=0$ holds since $\frac{\partial
{F}}{\partial \alpha}\varpropto\sin \frac{\alpha}{2}$. The optimal
setting of $\beta$ in Eq. (5.2), can be directly calculated from the
equation ${\partial {F}}/{\partial {\beta}}=0$ with
$F=\frac{1}{2}\{1+(1-\overline{{n^2_z}})\cos
(\frac{\pi}{4}-\frac{\beta}{2})+\frac{1}{2}(\overline{{n_z}^2}+\overline{n_z})\cos
\beta\}.$
\end{widetext}
\end{appendix}

\end{document}